UNIVERSITY OF PENNSYLVANIA

SCHOOL OF ENGINEERING AND APPLIED SCIENCE

# IN-SITU QUANTITATIVE MEASUREMENT OF ELECTRIC FIELDS IN ZINC OXIDE THIN FILMS USING ELECTROSTATIC FORCE MICROSCOPY

Jan Harloff

Philadelphia, Pennsylvania

(December, 1995)

A thesis presented to the Faculty of Engineering and Applied Science of the University of Pennsylvania in partial fulfillment of the requirements for the degree of Master of Science in Engineering for graduate work in Materials Science and Engineering.

______________________________________________

Dawn A. Bonnell (Supervisor)

______________________________________________

Dawn A. Bonnell (Graduate Group Chair)

## *Abstract:*

Zinc oxide (ZnO) is the most important material for the fabrication of modern varistors (*var*iable res*istors*). It is known that the highly nonlinear current-voltage relationship of ZnO varistors is due to effects taking place at the grain boundaries. Nevertheless, the exact mechanism of this process is not fully understood at present. For a more accurate investigation of this mechanism, techniques are required that allow a direct observation of local electric fields in varistor samples.

The objective of this study is to show that an atomic force microscope in a set-up as an electrostatic force microscope is capable of the in-situ observation and measurement of an applied electric field in samples of ZnO thin films. The technique of Surface Potential Imaging was used to investigate laterally applied electric fields in this type of sample for the first time. The local change of the electric field across the samples was monitored and quantified. It was observed here that the morphology of the surface is convoluted into the surface potential images and the magnitude of this effect was quantified by taking surface potential images without an applied electric field. For the given measurement conditions, a height difference of 80 nm in the topography image resulted in a voltage difference of roughly 66 mV in the surface potential image. A simple model was provided that attributes this observation to the surroundings of the surface atom closest to the imaging tip.

The estimated grain size of the samples is very small (approximately 140 nm). Even if it is assumed that the applied voltage drops entirely at the grain boundaries, the value of the drop per grain boundary is minute. The error introduced by the topography is one to two orders of magnitude larger than this drop. Therefore, the local voltage drop at a single grain boundary could not be observed. Instead, quantitative analysis was performed with large surface potential images (size of 80 μm x 80 μm). The effect of the topography was averaged out by taking mean values for the surface potential of a whole area. Thus, the voltage drop across the 80 μm sector was measured within the accuracy of the technique (roughly 16 mV). The local electric field was calculated from this voltage drop.

Samples with a much larger grain size than that of the ones used in this study are needed in order to measure the voltage drop at individual grain boundaries. The use of bicrystals or polycrystalline samples with a grain size large enough for the application of microcontacts could provide the necessary large voltage drop across a single boundary.

## *Acknowledgements:*

First of all I would like to thank Günter Schepker and Joyce Randolph, the hosts of the student exchange program between the Free University in Berlin, Germany, and the University of Pennsylvania. This program enabled me to come to the United States for one year, and I enjoyed this experience very much. I was very impressed by the Department of Materials Science and Engineering at the University of Pennsylvania, and I am deeply indebted to the whole LRSM for treating me very kindly and giving me an opportunity to learn so much. My special thanks go to my fellow first year graduate students, particularly to Tom Yeh, Roland Lee, and Daniel Polis. These three taught me the real meaning of the word "square", and the time spent together with them will stay among my most precious memories of my stay in America.

In the department I had the great fortune to join Prof. Dawn Bonnell's group, where I found a productive and inspiring working environment. Prof. Bonnell's dedication, understanding, and support enabled me to fulfill all the requirements for the degree of Master of Science in Engineering within the one year of my stay, for which I am very thankful. I could not have managed to write this thesis without the help of the whole group, particularly James Kiely, who introduced me to the atomic force microscope, Bryan Huey, who explained to me the method of Electrostatic Force Microscopy, and Dave Carroll, who was great in answering whatever question I was able to think of. I would furthermore like to thank V. Srikant, V. Sergo, and D.R. Clarke from the University of California, Santa Barbara, for providing the samples of the zinc oxide thin films.

Last, but not least, I am very grateful to my parents, Hans Joachim and Christiane Harloff. The love, care, and education that I received from them enabled me to pick up academic studies in the first place. Without their emotional and financial support I could not have come this far.

# *Contents:*

## Chapter 1: Introduction and Objectives

*In the interest of clearness, it appeared to me inevitable that I should repeat myself frequently, without paying the slightest attention to the elegance of the presentation. I adhered scrupulously to the precept of that brilliant theoretical physicist L. Boltzmann, according to whom matters of elegance ought to be left to the tailor and to the cobbler.*

> Albert Einstein
> "Relativity, The Special and the General Theory (A Clear Explanation that Anyone can understand)"
> Preface

### 1.1. Introduction:

A varistor is an example of a *smart material*. Smart materials are defined as materials that have the intrinsic ability to respond to their environment in a useful manner[1]. In the case of a varistor (whose name comes from *var*iable res*istor*) the response to environmental conditions lies in the highly nonlinear current-voltage (I-V) relationship. When exposed to a voltage higher than a certain breakdown value, a varistor loses most of its electrical resistance and current is conducted readily[2]. This characteristic is very helpful for surge protection so that varistors are widely utilized in protecting electric power lines and electronic components against dangerous voltage surges[3].

Of particular interest to modern surge protection are varistors made from *zinc oxide* (ZnO)[4,5]. The special importance of ZnO varistors is due to the fact that their nonlinear characteristics and the range of voltage and current over which the device can be used is far superior to those of SiC-based devices, the most popular surge protectors prior to the advent of the ZnO varistor[6]. The ZnO varistors were first developed in Japan by Matsuoka[7] and his research group in 1968 and commercialized in the following year. In the first decade after their invention various additives improving the electrical characteristics were discovered and the processing conditions were optimized. In the next decade, the microstructures and the physical properties of the grain boundaries were gradually identified. At that time applications grew in protection of electrical equipment and electronic components such as transistors and ICs against voltage surges. In 1988 almost 100 % of the lightning arresters produced in Japan were ZnO varistors[3].

The important effect producing the nonlinear I-V characteristic takes place at the grain boundaries. This is a very small structure and thus difficult to observe directly. Most of the models of varistor action were originally based on macroscopic observations of a bulk sample behavior. To clarify fundamental issues that remain the subject of discussion, techniques are required that allow a direct observation of the local electric field at the grain boundaries.

**1.2. Objectives:**

In this study, an atomic force microscope in a set-up as an electrostatic force microscope is used to perform surface potential measurements on two different samples of ZnO thin films that were prepared by laser ablation of polycrystalline ZnO targets (doped with 1 atom-% aluminum). The objective of the experiments is to show that the in-situ observation of an applied electric field in ZnO is possible and that quantitative values of the local surface potentials can be measured.

## *Chapter 2: Theoretical Background*

> ```
> If we do discover a complete theory, it should
> in time be understandable in broad principle to
> everyone, not just a few scientists. Then we
> shall all, philosophers, scientists, and just
> ordinary people, be able to take part in the
> discussion of the question of why it is that we
> and the universe exist. If we find the answer
> to that, it would be the ultimate triumph of
> human reason - for then we would know the mind
> of God.
> ```
>
> Stephen Hawking
> "A Brief History of Time - A Reader's Companion"
> Chapter 5

### 2.1. Introduction:

This chapter gives an overview about the topic of surge protection and the particular importance that ZnO varistors have in this respect. The focus is on the properties of grain boundaries and, in particular, the formation of potential barriers, which is a key issue for the nonlinear I-V characteristics of ZnO varistors.

### 2.2. Surge Protection:

Power systems are operated with a constant line voltage and supply power to a wide variety of equipment. The power levels range from a few watts to megawatts, and the voltage at which the energy is generated, transported, and distributed ranges from hundreds of volts to hundreds of kilovolts. In order to provide efficient and economic transport of the energy over long distances transmission and primary distribution of this power are made at high voltages, tens to hundreds of kilovolts. The final utilization is generally in the range of 120 V, typical residential, to less than one thousand, industrial, and a few thousand volts for large loads[5].

At all of these voltage and power levels, no matter how high, the equipment depends upon maintenance of a constant *operating voltage* because it has only a limited capability of withstanding voltages exceeding the normal level. Telecommunications and data-processing systems operate at lower voltages, and the increased use of solid-state devices makes them even more vulnerable to transient overvoltages[5].

Power systems are exposed to external influences that can couple energy into the circuits, causing a momentary overvoltage or over-current. Internal switching of loads in these systems can also create these momentary events, generally called *surges*. Another phenomenon, the discharge of electrostatic charges built upon the human body or objects, can also inject unwanted voltages or currents into the circuits. Surges have many consequences on equipment, ranging from no detectable effect to complete destruction. In general, electromechanical devices withstand voltage surges until a dielectric breakdown occurs, but electronic devices can have their operation upset before that failure occurs. The consequences of a breakdown depend on the particulars of the situation. Generally current surges occurring in circuit components offer a low impedance to the current. Thus low-impedance paths are established along which the current can flow[5].

There are theoretically three ways to ensure the survival or the undisturbed operation of electrical equipment[5]:

1) Eradication of the cause of surges. This is not realistic since it would imply the elimination of lightning, for example.

2) Building equipment immune to any level of surges, no matter how high. This is technically not a feasible solution.

3) Finding the best economic trade-off, in which moderate surge-withstanding capability is built into the equipment and the worst surges occurring in the environment are reduced by the application of suitable protective devices to a level the equipment can tolerate. This is the most reasonable concept.

Because the source of the surge is actually a current rather than a voltage, effective protection is more likely to be achieved by diverting this current than by attempting to block it. The ideal current diverter should offer a very low impedance to the surge currents. However, in a power system where voltage is present at all times, this low impedance would draw excessive current because of the power loss associated with that current. The diverter therefore has to meet two contradictory requirements: low impedance for a surge but high impedance for the power system voltage. This can be achieved by using *nonlinear devices*, which exhibit a high impedance at low voltages and a low impedance at high voltages[5].

The nonlinear conduction of a device known as *clamp* allows thousands of amperes to be diverted during the surge, but only milliamperes or less at the normal system voltage. Figure 2.1 schematically shows a power circuit which is subject to surge occurrences with a load L that the shunt-connected clamp D is expected to protect. A surge can be described as a current source that forces the flow of current through Z toward the parallel combination of L and D. Because of the nonlinear characteristic of conduction in D, its dynamic impedance is much lower than that of L so that, for practical purposes, all of the surge current passes through D. Furthermore, the low impedance of D results in a relatively low voltage developed across D, which is the voltage imposed by the load. Thus, the overvoltage protection of the load is ensured in spite of the threat of a current surge[5].

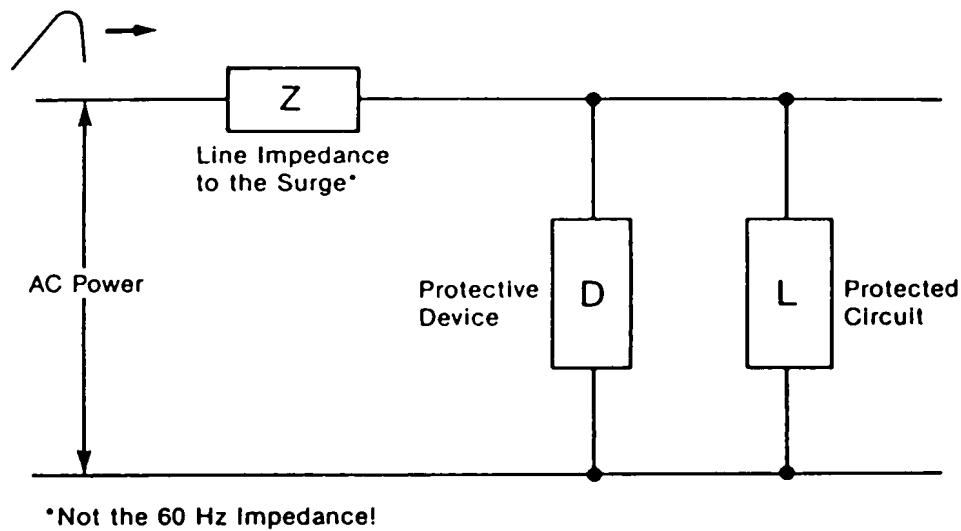

Fig. 2.1: Schematic elements of surge protection by current diversion (taken from Ref. 5).

Clamping devices pass from the high impedance to the low impedance state without abrupt change. The nonlinear I-V relationship can be described by the simple empirically defined equation

$$I = \left(\frac{V}{C}\right)^{\alpha} = K \times V^{\alpha}$$

(2.1)

where C and K are material constants[6]. The practical implication of Equation (2.1) is that, for a given discharge current, the voltage rise is lowered with increasing values of α. Therefore α is the figure of merit for the quality of a current diverter device. The significant advantage of zinc oxide varistors over the previously used silicon carbide varistors is their high value of α. Whereas SiC has a value[6] of α ≈ 5 the values for ZnO varistors[3,6,9] are α ≈ 30-100. Schematic I-V curves for different values of α are shown in Figure 2.2.

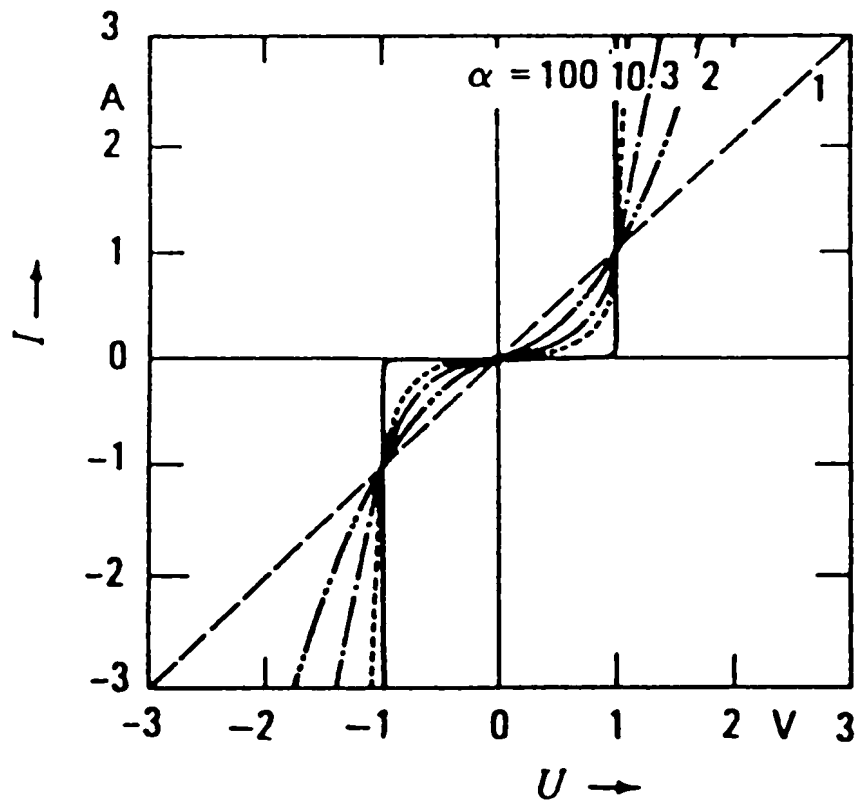

Fig. 2.2: Schematic I-V curves for different values of the nonlinear coefficient α (taken from Ref. 2).

## 2.3. Fundamental Characteristics of Zinc Oxide:

ZnO crystallizes in the hexagonal wurtzite structure in which the oxygen atoms are arranged in a hexagonal close-packed type of lattice with zinc atoms occupying half the tetrahedral sites (see Figure 2.3). The two types of atoms, Zn and O, are tetrahedrally coordinated to each other and are therefore equivalent in position. The mean lattice constants are a = 3.250 Å and c = 5.206 Å, depending slightly on stoichiometry deviation. The c/a-ratio of 1.60 is a little smaller than the ideal value of 1.633. The Zn-O distance is 1.992 Å parallel to the c-axis and 1.973 Å in the other three directions of the tetrahedral arrangement of nearest neighbors[10].

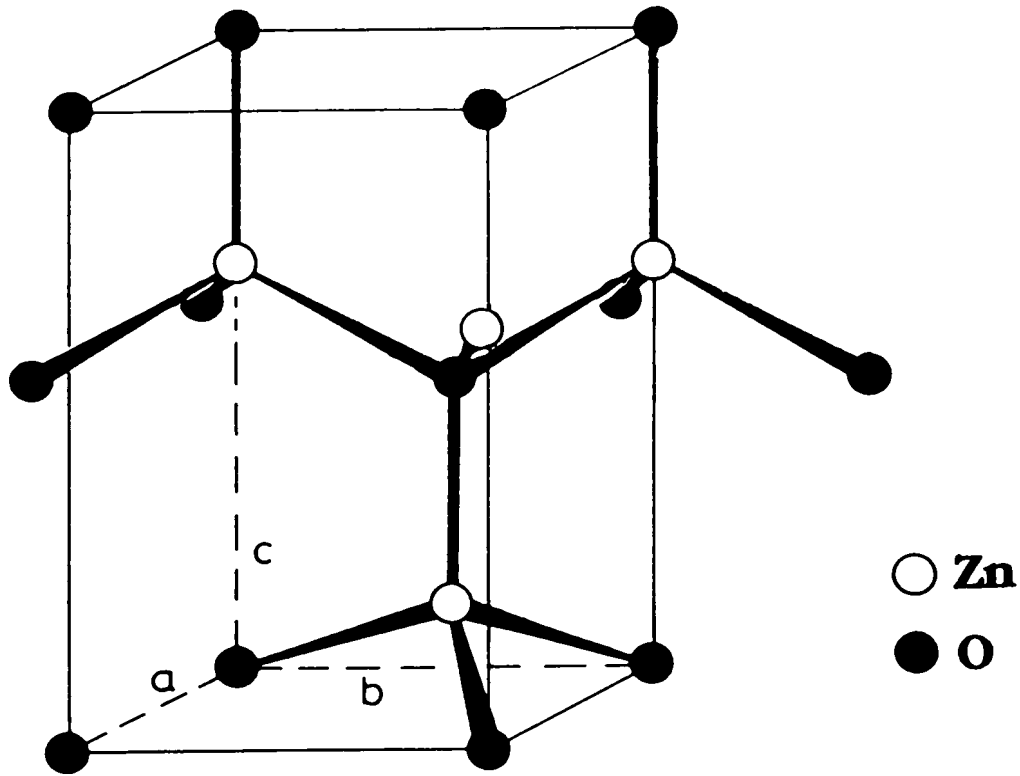

Fig. 2.3: Wurtzite structure of ZnO (taken from Ref. 11).

The specific gravity is 5.72 g×cm$^{-3}$, corresponding to 4.21×10$^{22}$ molecules per cubic centimeter[10]. Pure ZnO is an intrinsic semiconductor with a direct band gap (see Figure 2.4). The values for the minimal band gap stated in the literature range from 3.1 eV[10] to 3.3 eV[6]. The Zn 3d-band is energetically situated just below the valence band[10].

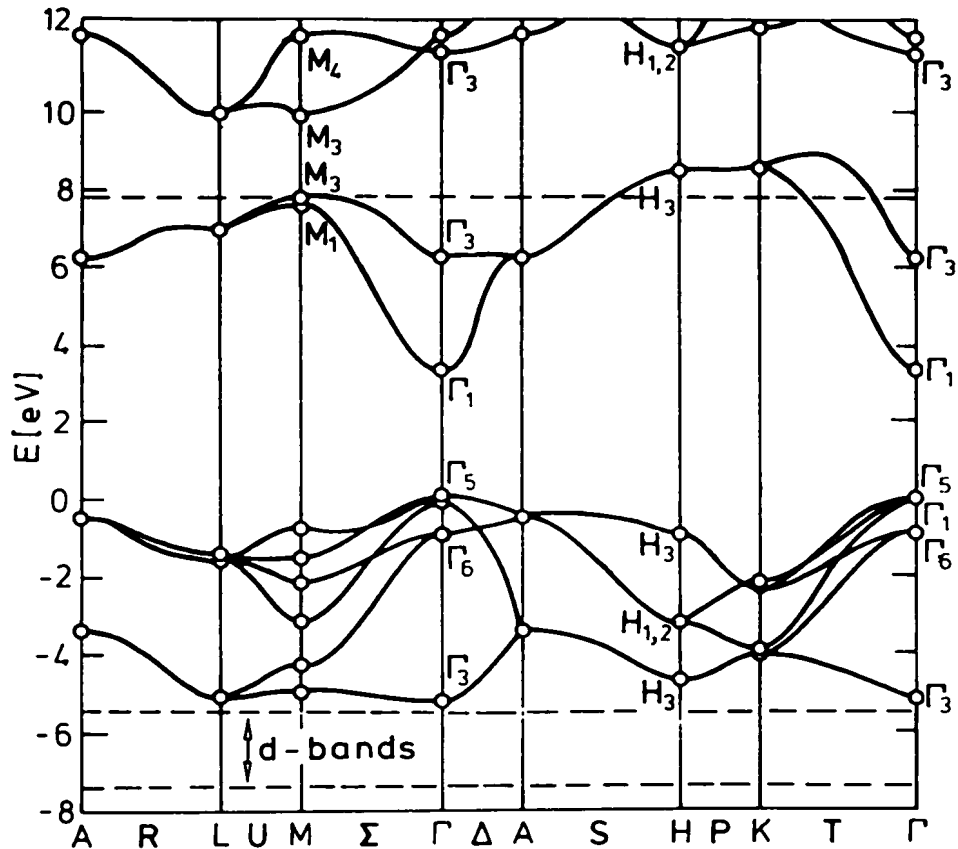

Fig. 2.4: Energy bands of ZnO (taken from Ref. 10).

The ZnO structure is relatively open, with all of the octahedral and half of the tetrahedral sites empty. It is, therefore, relatively easy to incorporate external dopants into the ZnO lattice. The open structure also influences the nature of defects and the mechanism of diffusion[6]. Single crystals of ZnO exhibit n-type conductivity, i.e. ZnO is not an intrinsic semiconductor, because there is excess zinc acting as a donor. The zinc excess results in a nonstoichiometric compound $Zn_{1+\delta}O$ and structural disorder[12]. Two types of defects must be considered for the defect chemistry of ZnO: *extrinsic defects* due to additives and *intrinsic defects* due to the inherent nonstoichiometry of ZnO. The dominant intrinsic defects have long been a subject of controversial discussion in the literature[13]. One double acceptor site, which is the *zinc vacancy*, has generally been accepted[13].

Regarding native donor-like defects, there were two competing opinions. One favored the interstitial zinc site and the other favored the oxygen vacancy. It now seems to be established that the *oxygen vacancies* are the major donor like defect[14, 15, 16].

### 2.4. Nonlinear I-V Characteristics of a ZnO Varistor:

The most important property of a ZnO varistor is its highly nonlinear I-V characteristic. The I-V curve of a ZnO varistor can be divided into three important regions (see Figure 2.5)[6]:

1) Low-current linear region, below the *threshold voltage* (typically a voltage at hundreds of μA/cm$^2$). This region is defined as the *prebreakdown region* and the I-V characteristic is linear. The AC current is about two orders of magnitude higher than the DC current for a given operating voltage. The difference can be attributed to the contribution of the dielectric loss upon application of an AC voltage. The I-V characteristic in this region is highly temperature dependent[3] and determined by the impedance of the grain boundaries in the ZnO microstructure.

2) Intermediate nonlinear region, between the threshold voltage and a voltage at a current of about $10^2$-$10^3$ A/cm$^2$. This *nonlinear region* of the intermediate current is the most important feature of a ZnO varistor. Here the device conducts an increasingly large amount of current for a small increase in voltage. The non-linear region can extend over six to seven orders of magnitude of current. The I-V characteristic in this region is almost independent of temperature[3].

3) High-current linear region, above approximately $10^2$-$10^3$ A/cm². In this region, which is known as the *upturn region*, the I-V characteristics are linear again, similar to that in the low-current region, with the voltage rising faster with current than in the nonlinear region. This region is controlled by the impedance of the grains in the ZnO microstructure.

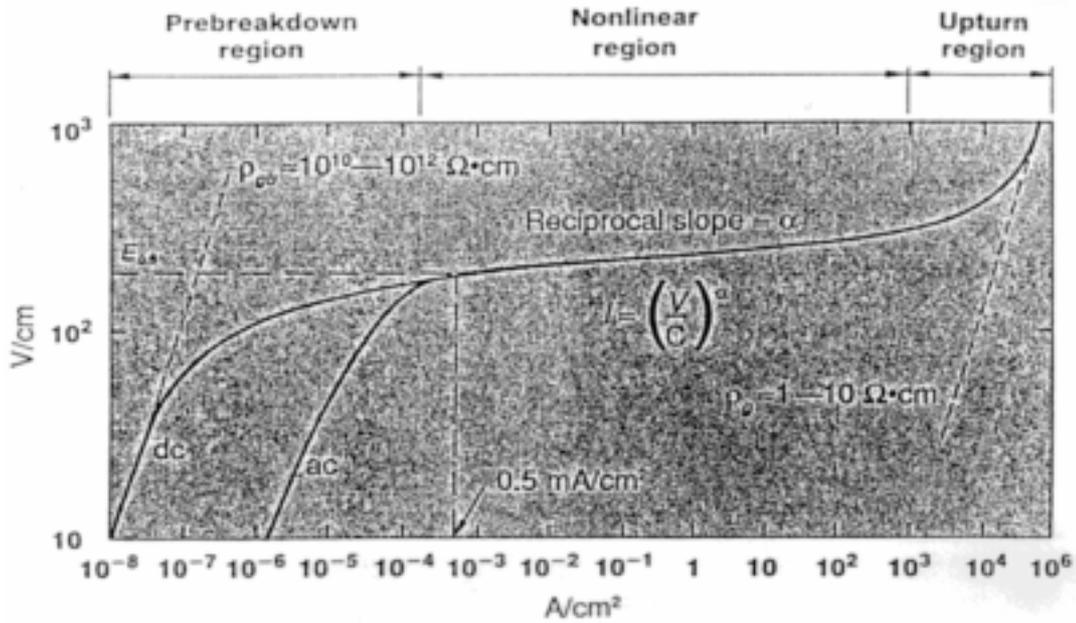

Fig. 2.5:  Typical I-V curve of a ZnO varistor over a wide range of current density and electric field (taken from Ref. 6).

The most important parameter of the ZnO varistor is the nonlinear coefficient $\alpha$, which is the reciprocal of the slope of the I-V curve in the nonlinear region, and is defined by[6]:

$$\alpha = \frac{d \ln I}{d \ln V}$$

(2.2)

The greater the value of α, the better the device; in a perfect varistor α would go to infinity. It is important to note, however, that the current ranges must be clearly stated whenever the α value is cited to claim the quality of a device. It is usually cited for current ranges lying between 0.1 and 100 mA. For high-current applications, e.g. 1 kA, choosing α values from the literature must be done with care. Moreover, the value of α is affected by the temperature and pressure to which the device is subjected during application. With both temperature and pressure the α value decreases, clearly indicating the need to control external environments during application[6].

The ZnO varistor is characterized by a voltage which marks the transition from linear to nonlinear mode. The voltage at the onset of this nonlinearity, just above the "knee" of the I-V curve (see Figure 2.5), is the threshold voltage which determines the voltage rating of the device. This is also known as nonlinear or "turn-on" voltage. Because of a lack of sharpness of the transition in the I-V curve, the exact location of this voltage is difficult to determine in most varistors. However, from Equation (2.1) the threshold voltage can be defined as[6]:

$$C = \frac{V}{I^{1/\alpha}} = V \text{ at } 1 \text{ mA}$$

(2.3)

Using this definition, the varistor threshold voltage has often been described in the literature as the voltage observed at 1 mA ($V_{1 \text{ mA}}$). Other authors have used the voltage at 10 mA ($V_{10 \text{ mA}}$) as the reference voltage. None of these definitions, however, takes into account the effect of the geometry of the device. This can be done by using the normalized values of voltage and current. This *normal voltage* has been defined as the electrical field, $E_{0.5}$, measured at the current density of 0.5 mA/cm² (see Figure 2.5). Experience has shown that, for most varistors, $E_{0.5}$ values are close to the onset of nonlinearity[6].

## 2.5. Fabrication of Zinc Oxide Varistors:

ZnO varistors are highly complex, multicomponent, polycrystalline oxide ceramics whose electrical behavior depends both on the microstructure of the device and on detailed processes occurring at the ZnO grain boundaries. The primary constituent of such a varistor is obviously ZnO, typically 90 % or more. In addition the varistor contains smaller amounts of a number of other metal oxide constituents. A typical composition[4,5] contains 97 mol-% ZnO, 1 mol-% $Sb_2O_3$, and 0.5 mol-% each of $Bi_2O_3$, CoO, MnO, and $Cr_2O_3$. Fabrication of ZnO varistors follows *standard ceramic techniques* (see Figure 2.6)[4,5].

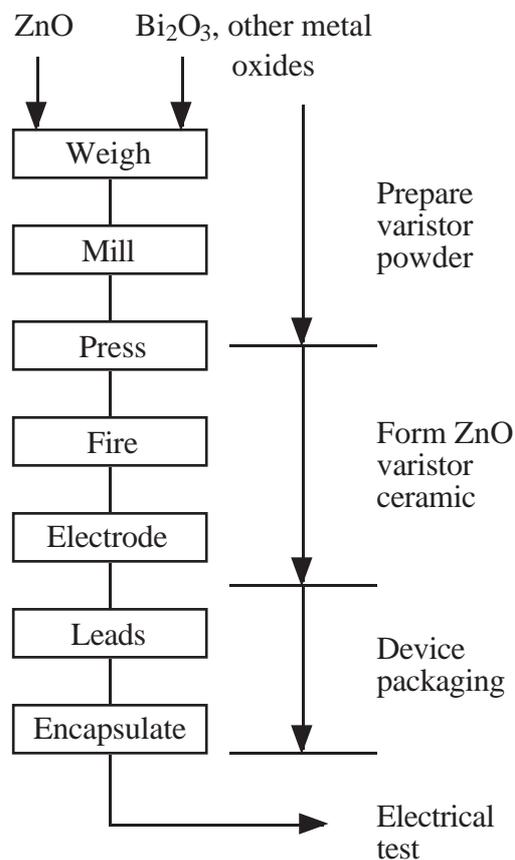

Fig. 2.6: Simplified flow diagram for the fabrication of ZnO varistors (after Ref. 5).

The components are mixed, for example by milling in a ballmill and spray drying afterwards. The mixed powder is dried and pressed to the desired shape. The resulting pellets are sintered at high temperature, typically 1000 to 1400 °C. Electrodes are then attached to the sintered devices, often in form of a fired silver contact. The device behavior is not affected by the electrode configuration or its basic composition. Leads are generally attached by solder, and the device may be encapsulated in a polymeric material. The finished product is usually electrically tested to meet required specifications for performance characteristics[4,5].

The basic behavior of ZnO varistors is a consequence of the addition of *varistor-forming* ingredients to ZnO. These are heavy elements such as Bi, Pr, Ba, and Nd with large ionic radii[9]. The most pronounced varistor-forming effect is produced by Bi and Pr[3]. The usual varistor-forming additive of choice for commercial applications is $Bi_2O_3$[4,5,6]. $Bi_2O_3$ creates potential barriers at the grain boundaries by supplying ions which reside in the region of the ZnO grains near the boundaries. The nonlinear property appears by adding at least 0.1 mol-% of $Bi_2O_3$. However, the $\alpha$ values achieved in this way never exceed 10[3]. By adding *varistor performance ingredients* the nonlinearity is dramatically improved. These are generally transition metal elements, such as Co, Mn, and Ni[9]. The $\alpha$ values achieved in this way not only exceed 10, but reach values as high as 40[3].

The addition of $Sb_2O_3$ leads to the formation of $Zn_7Sb_2O_{12}$ spinel-type crystallites at the grain boundaries during sintering. The $Zn_7Sb_2O_{12}$ particles at the grain boundaries hinder the ion transfer which results in suppression of grain growth. Small grain sizes further improve the nonlinear property. Furthermore, the I-V characteristics become stable against electrical stresses. This is important in the practical applications of these materials[3].

As mentioned above, the upturn region of the I-V curve is grain-resistivity controlled. Both $Al^{3+}$ and $Ga^{3+}$ have been shown to decrease the grain resistivity and increase the high-current nonlinearity of the varistor[6]. Using all of the mechanisms stated above, α values as high as 100 can be achieved for ZnO varistors[3,6,9]. As a summary it can be said that commercially manufactured ZnO varistors usually contain the *four basic additives* $Bi_2O_3$, $CoO$, $MnO$, and $Sb_2O_3$, along with some additives to control grain size, grain resistivity, and ZnO stability[3].

## 2.6. Microstructure of Zinc Oxide Varistors:

The basic building block of the ZnO varistor microstructure is the ZnO grain formed as a result of sintering. The size of ZnO grains is usually 5-30 μm in commercial varistors[3,4,5,6,9] and depends on the material composition, sintering temperature, and sintering time. The grains are embedded in a three-dimensional open network of an intergranular, amorphous, Bi-rich phase, which usually contains several crystalline components (see Figure 2.7)[18,19]:

1) Different polymorphs of crystalline $Bi_2O_3$[2,3,6,19,21,22]: α-, β-, γ-, and δ-$Bi_2O_3$. The particular polymorph of $Bi_2O_3$ that forms is a direct function of cooling rates and heat treatment conditions[19].

2) If present in large amounts, several of the metal oxide additives react with ZnO to form a phase with the *spinel* crystal structure[23]. The most important one is $Sb_2O_3$ (forming $Zn_7Sb_2O_{12}$)[3,19], other examples are $Al_2O_3$ (forming $Zn_7Al_2O_4$), $TiO_2$ (forming $Zn_2TiO_4$), and $Nb_2O_5$ (forming $Zn_3Nb_2O_8$)[23].

3) A *pyrochlore-type phase*, $Zn_2Bi_3Sb_3O_{14}$[3,19].

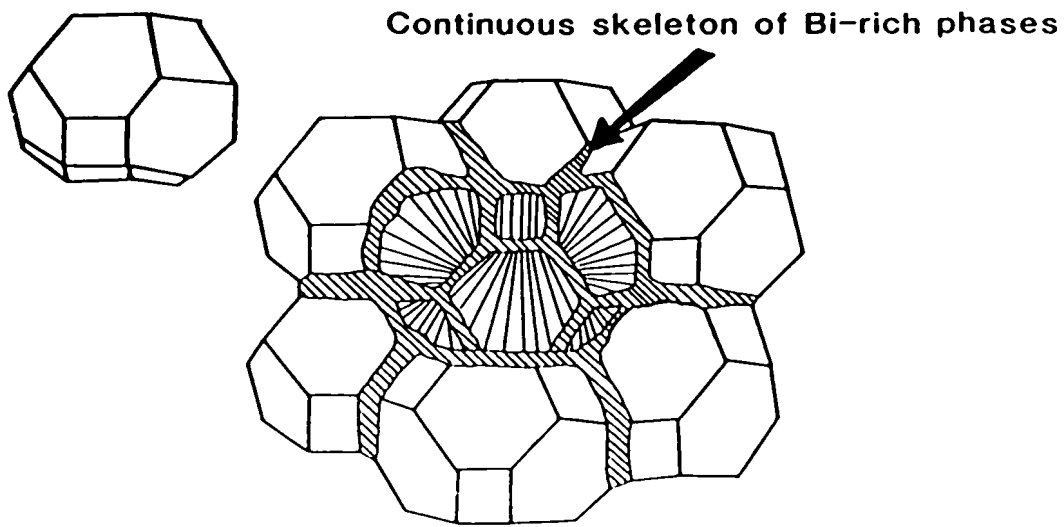

Fig. 2.7: Three dimensional grain structure of ZnO varistor material showing the continuous skeleton of Bi-rich phase (taken from Ref. 20).

The microstructures at the grain boundaries are quite complicated. They can be roughly classified into three types of structures (see Figure 2.8)[3,19,24]:

1) *Type I* grain boundaries have a relatively thick (more than 100 nm) $Bi_2O_3$-rich amorphous intergranular layer. This type of grain boundary may contain smaller crystallites of secondary phases.

2) *Type II* grain boundaries have a thin (1 to 100 nm) $Bi_2O_3$-rich amorphous intergranular layer.

3) *Type III* grain boundaries are characterized by a direct contact of the grains, without an intergranular layer in between. Bi, Co, and an excess amount of oxygen ions can be detected in the interfacial region of these grain boundaries to a thickness of several nanometers.

The grain boundaries of type II and III are the important structures in the material, giving rise to the varistor effect[21].

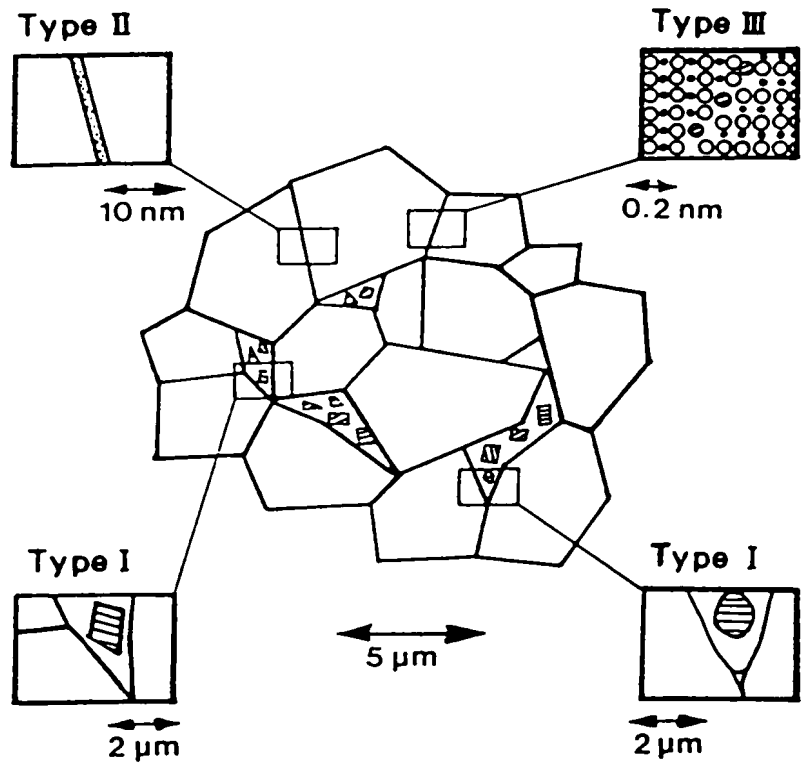

Fig. 2.8: Schematic structure of grain boundaries in ZnO varistors (taken from Ref. 21).

In the sintering process of a standard composition for a multicomponent varistor, the $Sb_2O_3$ reacts with both ZnO and $Bi_2O_3$ above 700° C and forms the spinel- and pyrochlore-type phases $Zn_7Sb_2O_{12}$ and $Zn_2Bi_3Sb_3O_{14}$. During sintering the $Zn_2Bi_3Sb_3O_{14}$ forms a $Bi_2O_3$-rich liquid phase and $Zn_7Sb_2O_{12}$, which precipitates at the grain boundaries where it hinders ion transfer. As a result the grain growth during sintering is suppressed. The liquid is likely to gather at the triple points of the ZnO. During cooling the liquid phase changes to amorphous $Bi_2O_3$-rich intergranular layers. Therefore the type I grain boundaries are likely to be found at the packing holes in the sintered body. The $Bi_2O_3$-rich intergranular layer becomes thinner as it approaches the contact points of the particles, so the grain boundaries become type II. At the contact points, finally, no intergranular layer can be observed, so that the grain boundaries are of type III (see Figure 2.9)[3].

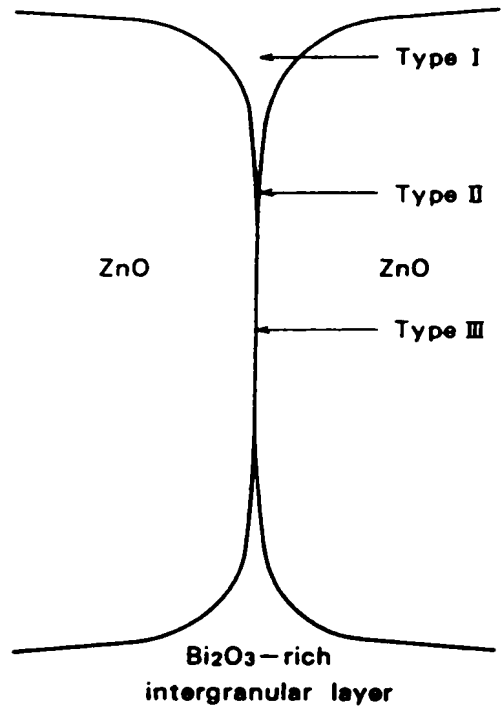

Fig. 2.9:   Development of the grain boundaries during sintering (taken from Ref. 3).

$Bi_2O_3$ easily vaporizes above 1400 °C, and even at 1200 °C, a usual sintering temperature, $Bi_2O_3$ vaporizes from the surface of the device; therefore, the amount of $Bi_2O_3$ gradually decreases from the sintered mass. Furthermore, precipitation of oxides from the $Bi_2O_3$ liquid phase occurs during cooling because the $Bi_2O_3$ liquid phase contains a number of ions, such as Zn, Co, Mn, and Sb. A large amount of ZnO can dissolve in the $Bi_2O_3$ liquid phase when $Sb_2O_3$ is present. Hence, precipitation of ZnO occurs at the grain boundaries during cooling (see Figure 2.10). As a result of those two mechanisms the amount of the $Bi_2O_3$ phase diminishes during the sintering process, leaving precipitated ZnO and some ions of Bi, Co, Mn, and Sb at the grain boundaries.

The wettability of ZnO grains by the $Bi_2O_3$ liquid phase is not very good; therefore, when the amount of $Bi_2O_3$ is diminished, the ZnO grains are not completely surrounded by the $Bi_2O_3$ liquid phase. However, the diffusion velocity at a grain boundary is usually higher than that in the bulk by one or two orders of magnitude. As a result, ions of Bi, Co, Mn, and Sb diffuse into the grain boundaries with ease. The mechanisms just described account for the three different types of grain boundaries. The relative amounts of the types of grain boundaries can be quite different from material to material, and depend on both composition (especially $Bi_2O_3$ content) and sintering conditions[3].

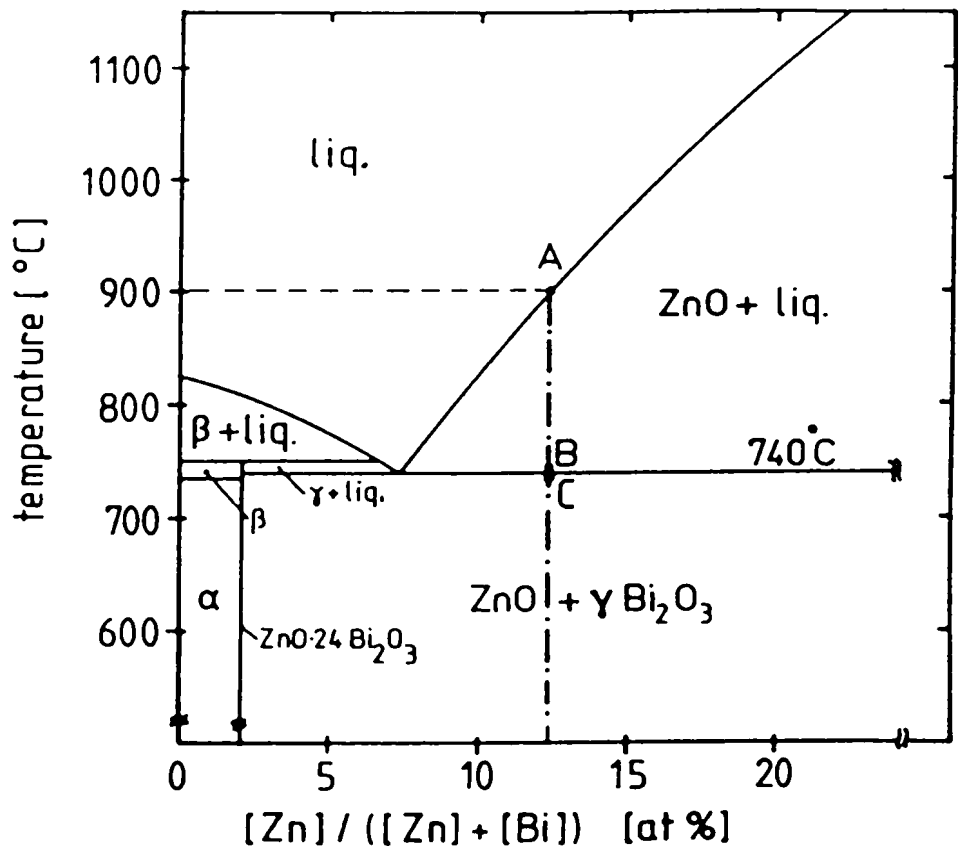

Fig. 2.10: Phase diagram of the system $ZnO-Bi_2O_3$. During cooling (at the point A) solid ZnO starts to precipitate from the liquid (taken from Ref. 25).

## 2.7. Block Model:

During the sintering process, various chemical elements are distributed in such a way in the microstructure that the near-grain-boundary region becomes highly resistive ($\rho_{gb} \approx 10^{10}$-$10^{12}$ $\Omega \times cm$)[3,6] and the grain interior becomes highly conductive ($\rho_g \approx 0.1$-$10$ $\Omega \times cm$)[3,4,5,6]. A sharp drop in resistivity from grain boundary to grain occurs within a distance of approximately 50 to 100 nm, known as the *depletion layer*[3,4,5,6]. Thus, at each grain boundary, there exists a depletion layer on both sides of the grain boundary extending into the adjacent grains. The varistor action arises as a result of the presence of this depletion layer within the grains. Since this region is depleted of electrons, a voltage drop appears across the grain boundary upon application of an external voltage[4,5].

To analyze the macroscopic behavior of a varistor it is useful to represent the microstructure by the *block model*[4,5]. This model presumes the device to be assembled of conducting ZnO cubes of size, d, which is given by the average grain size (see Figure 2.11). The cubes are separated from each other by the insulating depletion layers of thickness, t. The *breakdown voltage* per intergranular barrier, $V_{gb}$, can be calculated by multiplying the cube size with the macroscopic average breakdown field, $F_B$[4,5]:

$$V_{gb} = F_B \times d$$

(2.4)

The value calculated in this way is lower than the true breakdown voltage per grain boundary, because the current always seeks the easiest path, i.e. the path with fewest barriers between the electrodes. Hence, the number of grains for the current path is lower than the average number of grains between the electrodes. It is an important and significant feature of the behavior of ZnO varistors that breakdown voltages of 2 to 4 V per grain boundary[3,4,5,6,17,26] are observed for a wide variety of ZnO varistor materials. Substantial variations in device processing and composition have relatively minor effects upon $V_{gb}$[3].

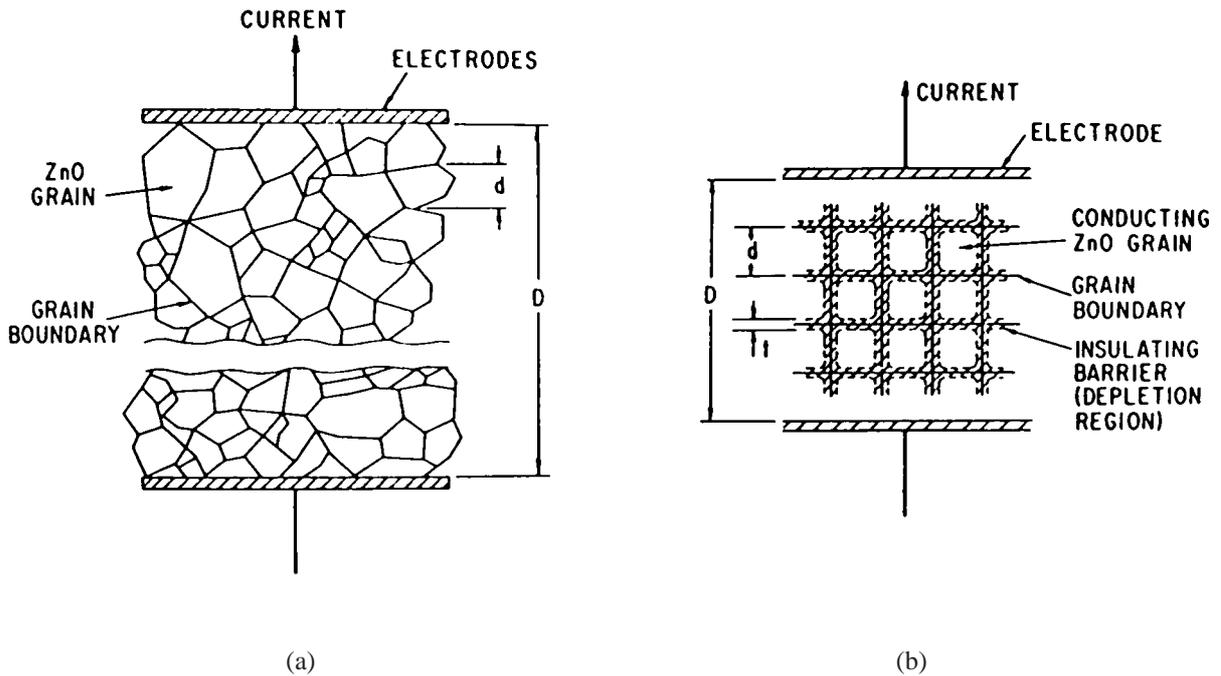

Fig. 2.11: (a) Schematic depiction of the microstructure of a ZnO varistor. The number of grain boundaries between the electrodes in the macroscopic sample of length, D, is a function of the average grain size, d. (b) Block model (taken from Ref. 5).

The block model shows that the electrical characteristics of ZnO varistors are related to the bulk of the material, i.e. the device is inherently a multijunction with varistor action shared between the various ZnO grain boundaries. This implies that tailoring the macroscopic device breakdown voltage, $V_B$, is simply a matter of fabricating a varistor with the appropriate number, n, of grains in series between the electrodes. Thus, to achieve a given breakdown voltage, one can change the varistor thickness, D, for fixed grain size, d, or, alternatively, one can vary the grain size to increase the number of barriers, n, keeping the device thickness constant. In either case the breakdown voltage is given by[4,5]:

$$V_B = n \times V_{GB} = \frac{D \times V_{GB}}{d}$$

(2.5)

Typical varistor values[4,5] for protection of equipment on 120 V AC power lines are $V_B$ = 200 V, d = 20 µm, D = 1.6 mm, and n = 80.

### 2.8. Potential Barriers at Grain Boundaries:

The high resistivity in the vicinity of grain boundaries in ZnO varistors is a consequence of *potential barriers* in that region. To some degree such grain boundary potential barriers always exist in polycrystalline semiconductors, since the chemical potentials of grain boundary regions are shifted from the bulk value by the lack of perfect periodicity. A more pronounced change of the chemical potential can be created by vacancies, impurities, or other phases in the boundary[27,28].

The number and nature of vacancies and impurities at grain boundaries are the result of *segregation*[29]. The major driving forces leading to segregation in ceramics are the reduction in the *elastic energy* and the *space charge regions* associated with the grain boundaries. When solutes have a size misfit with the matrix, their segregation to the inherently distorted grain boundary region provides a partial relaxation of the elastic energy[30].

The existence of space charge regions near grain boundaries in an ionic solid is due to the fact that the formation energy of cation and anion vacancies is different. Since a grain boundary acts as a source and sink of defects, there will be a larger concentration of the defect with the lower formation energy in its vicinity. The defects usually carry net charges, therefore constraints of charge neutrality will tend to maintain equal concentrations of the two types of defects in the bulk of the material. Between the interface and the neutral bulk, a space charge region will exist in which one type of vacancy predominates. The excess charge in this region will be compensated by an excess of the opposite charge in the interface[31,32].

In the particular case of ZnO varistors, the nonlinear I-V characteristic is a consequence of the fact that the grain boundary potential barriers are very large (see Figure 2.12)[28]; the height, $\phi$, is approximately 0.6 to 0.8 eV[2,4,5,6,9].

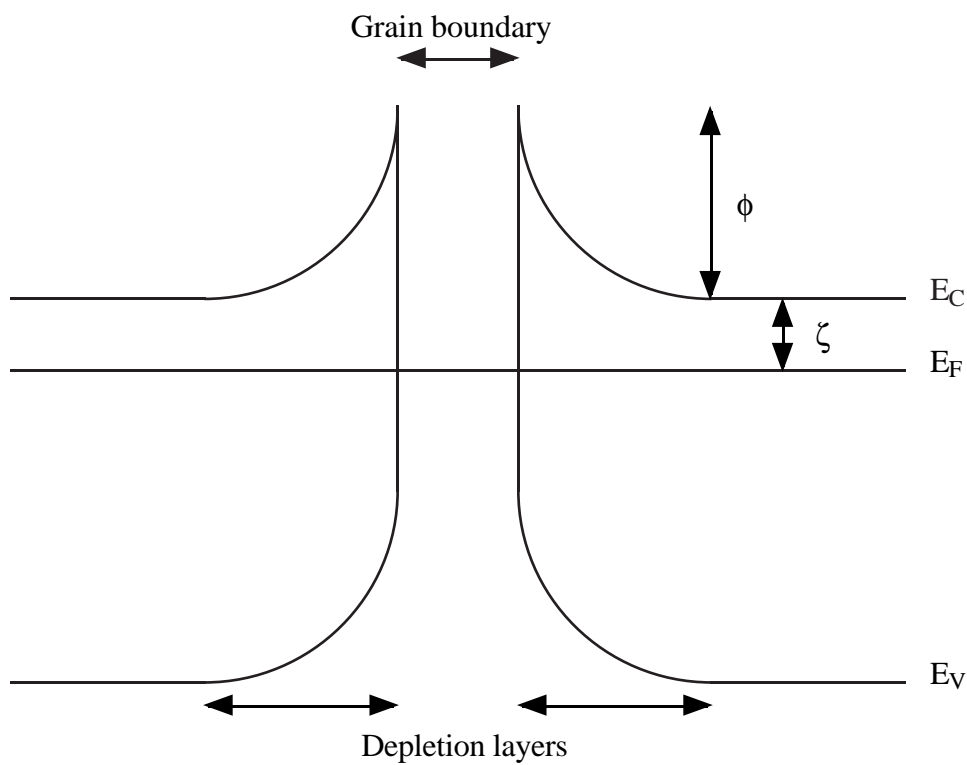

Fig. 2.12: Grain-boundary band diagram for a ZnO varistor. Band bending takes place in the depletion layers, at the sides of the grain boundary (after Ref. 33).

The mechanism leading to the formation of these very high potential barriers was the subject of a long and controversial discussion in the literature, and many models have been proposed. Every model that tries to successfully explain the formation of the potential barriers and the electrical behavior of ZnO varistors has to take several important experimental observations into account:

1) The addition of a varistor-forming ingredient (usually Bi) to the ZnO is necessary in order to produce the varistor characteristic[3,4,5,6,9].

2) The annealing process of the ceramic is extremely important:
   a) The presence of oxygen in the annealing atmosphere is necessary for the formation of the varistor characteristic[26].
   b) The cooling rate after the annealing treatment has to be moderate. Water quenched specimens do not exhibit varistor characteristics[21,34].

3) The breakdown voltage per grain boundary has a fixed value of approximately 3 V. This value is insensitive to large variations in the chemistry of the varistor ceramic[2,3].

## **2.9. Schottky Barrier Model:**

Most of the models for ZnO varistors describe the depletion layer as a *Schottky barrier*[13,15,16,33,35,36]. This name is usually used for the region that forms when a semiconductor is brought in contact with a metal. As an example one can consider the case for a metal and a n-type semiconductor. For the infinitely separated materials the Fermi level in the metal ($E_{FM}$) is less than the Fermi level in the semiconductor ($E_{FS}$); therefore, the average energy of electrons in the semiconductor is greater than that of those in the metal. When the materials are brought into perfect contact, the difference in the averaged electron energy transfers electrons from the semiconductor to the metal until the average electron energies are equal. This transfer of electrons from the n-type semiconductor to the metal leaves the surface of the semiconductor depleted of electrons and leaves behind some positive donor ions. This zone is therefore called the depletion layer. Having received electrons the metal is negatively charged with respect to the semiconductor. Figure 2.13 illustrates that a Schottky barrier actually corresponds to an energy barrier between the metal and the semiconductor[37].

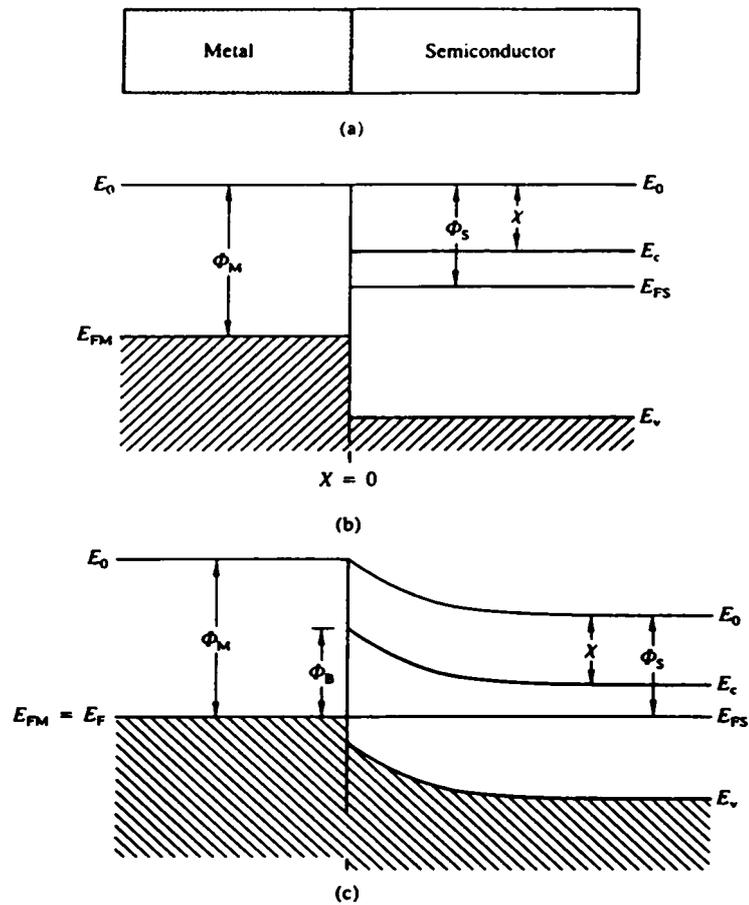

Fig. 2.13: Schematic representation of a Schottky barrier:
(a) physical contact of metal and semiconductor.
(b) Energy diagram for isolated metal and semiconductor.
(c) Energy diagram for the metal and semiconductor in contact (taken from Ref. 37).

A free electron with an energy of $E = E_{FM}$ moving from the metal to the semiconductor is confronted with a potential barrier of $\Phi_B$. A free electron moving from the semiconductor bulk at $E = E_C$ to the metal sees a potential barrier that is the difference in the two original Fermi energy levels before contact[37]. Transferring this model to varistors, the depletion layers at the surface of two ZnO grains in contact to each other can be described as two Schottky barriers connected back-to-back (see Figure 2.14)[13,15,16,33,35,36].

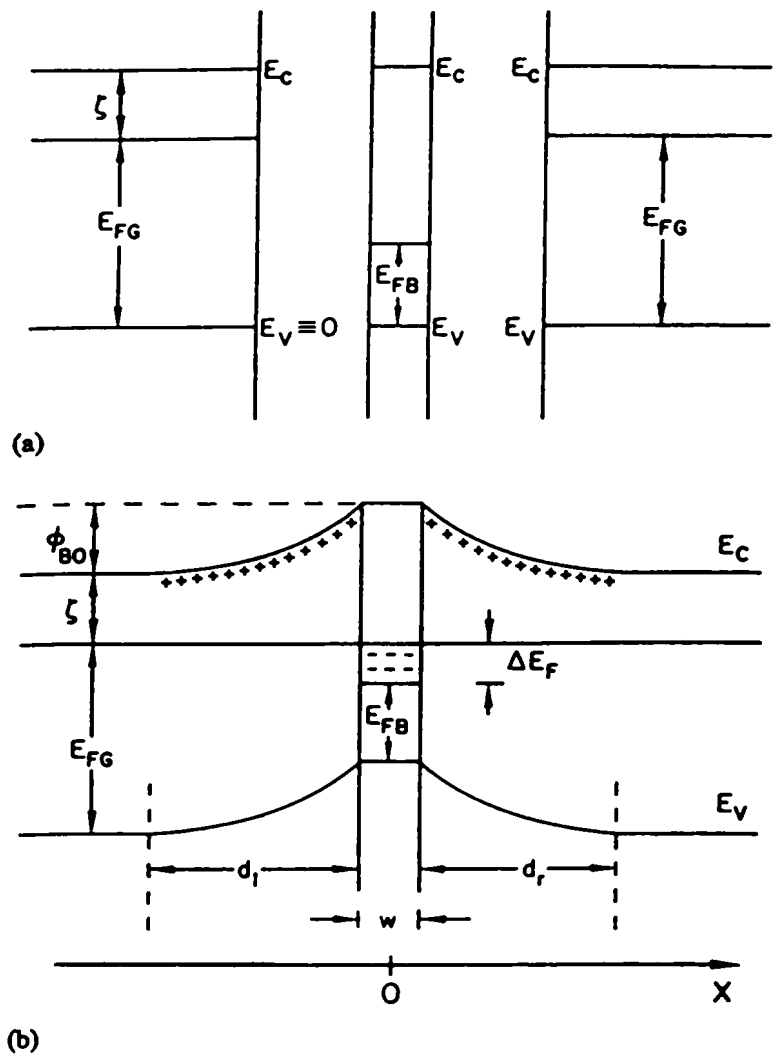

Fig. 2.14: Energy-band diagram for two grains in a ZnO varistor: (a) The two grains and the region of their boundary schematically as they would be if they were three isolated materials. (b) Band bending in the grains which occurs when the materials in (a) are joined at zero applied bias voltage (taken from Ref. 27).

The formation of the Schottky barriers is due to the trapping of electrons at defect states which are present in the region close to the grain boundary[27]. The defect states are connected to the doping ions (particularly Bi) which have a much higher concentration in the vicinity of the grain boundaries (as was described in detail in chapter 2.6, a moderate cooling rate after the sintering leads to an enrichment of the additives at the interfaces). Furthermore, the presence of oxygen during the sintering and cooling seems to be necessary since the defect states created by the Bi change their behavior in a reducing environment, making them unfit as trap states for electrons[26].

## 2.10. Hole-Induced Breakdown Model:

Based on the Schottky barrier description of the grain boundaries many models were proposed to explain and describe the nonlinear I-V characteristics of ZnO varistors. The issue is still not resolved, but most authors favor the *hole-induced breakdown model*, since its agreement with experiment is excellent[9,13,15,38]. According to this model, electrons transfer over the potential barriers by *thermionic emission* at low voltages. At a critical voltage (which corresponds to the breakdown voltage), electrons transferred over the grain interface have sufficient energy to create minority carriers by *impact ionization*[9]. The generated holes are trapped at the interface for a short time, after which they recombine with electrons trapped at interface states, which reduces the barrier height (see Figure 2.15). The accumulation of holes on one side of the junction furthermore decreases the width of the Schottky barrier, making it thin enough to provide *tunneling* of electrons across the barrier[15]. These two effects lead to the extreme drop in impedance at the threshold voltage.

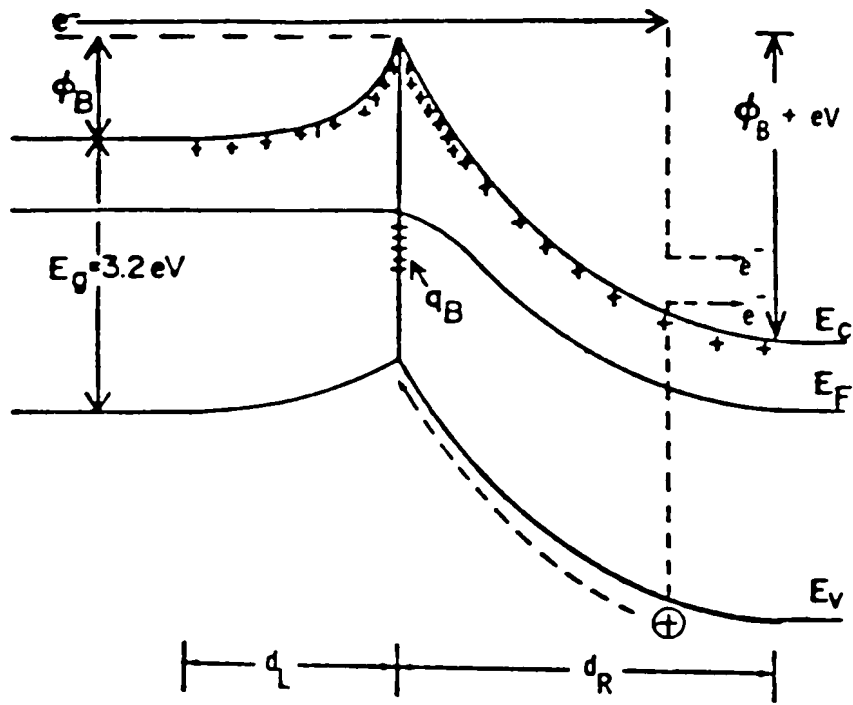

Fig. 2.15: Energy-band diagram for two grains in a ZnO varistor under an applied voltage corresponding to the breakdown voltage. Electrons crossing the boundary create holes that later recombine with the electrons trapped in the interface states (taken from Ref. 38).

## 2.11. Summary:

This chapter provided an overview about important issues regarding ZnO varistors. It was shown that several points regarding the nonlinear I-V characteristics are still subject of discussion in the literature. To solve some of these issues techniques that allow a direct observation of the electric field in ZnO varistors would be helpful. Surface potential measurements performed with an atomic force microscope are capable of providing such information.

## *Chapter 3: Surface Potential Measurements Using Force Microscopy*

THE FASHION WAS TO BLAME IT ON "TECHNOLOGY," BUT "TECHNOLOGY" IS THE TRUNK OF THE TREE, NOT THE ROOTS. THE ROOTS ARE RATIONALISM, AND I WOULD DEFINE THAT WORD SO: "RATIONALISM IS THE IDEA WE CAN EVER UNDER-STAND ANYTHING ABOUT THE STATE OF BEING."
... [NOW] WE MAY BE BEGINNING TO ACCEPT ... A DIFFERENT DEFINITION OF EXISTENCE. THE IDEA THAT WE CAN NEVER UNDERSTAND ANYTHING ABOUT THE STATE OF BEING.

> Glen Bateman from "The Stand (The complete & uncut edition)"
> by Stephen King
> Chapter 52

### **3.1. Introduction:**

This chapter introduces the concept of Force Microscopy (FM), which is a technique used for measuring forces between the surface of a sample and a tip on a length scale of $10^{-11}$ to $10^{-7}$ m. A force microscope basically consists of a sensor that responds to a force, a cantilever beam with a tip (often made out of silicon[39]) attached to it, and a detector that measures the sensor's response. A sample can be scanned underneath the tip to create a force map or image of the sample's surface[40,41,42]. The advantage of FM compared to Scanning Tunneling Microscopy (STM) is that neither the tip nor the sample need to be conductive. The disadvantage is that the resolution is not as good as that of STM. There are several forces that can be used for FM, for example magnetic forces.

The focus here will be on Atomic Force Microscopy (AFM) and Surface Potential Imaging (SPI) which is a particular form of Electrostatic Force Microscopy (EFM)[43]. SPI can be used to quantitatively measure the distribution of an electric field in a sample, comparable to potentiometry performed with a scanning tunneling microscope[44,45,46].

### 3.2. Atomic Force Microscopy:

An atomic force microscope is equipped with a tip at the end of a cantilever whose position in the x-, y-, and z-direction can be controlled very accurately with elements made from piezoelectric ceramics (abbreviated as "piezos"). The x- and y- piezos are used to scan the tip across the surface of a sample, the z-piezo is used to control the distance between the tip and the sample. The force on the cantilever tip as a function of the distance between the tip and a sample surface can be presented as a *force curve* (see Figure 3.1)[40].

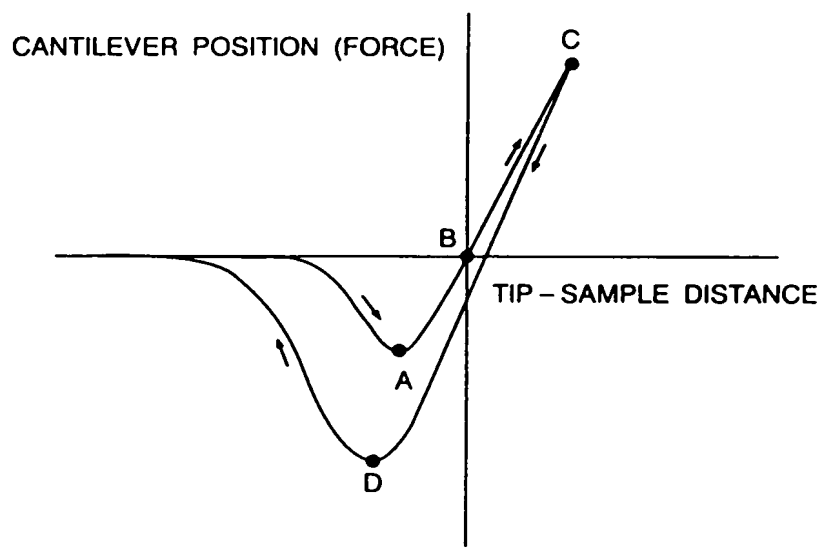

Fig. 3.1: Typical force curve showing the force on a cantilever tip as a function of tip-sample distance. Following the arrows, the tip is brought closer to the sample, into contact with it, and is removed from it again (taken from Ref. 40).

As long as the sample is far away from the cantilever (large negative values on the y-axis in Figure 3.1) the cantilever feels no force and is in its rest position (a value of zero on the x-axis). Upon decreasing distance between the sample and the tip, the cantilever bends towards the sample due to attractive surface forces. Point A in Figure 3.1 corresponds to the maximum attractive force, moving the cantilever even closer to the sample pushes the cantilever back through its original rest position in point B. The motion is continued until a predetermined load is applied to the sample by the cantilever in point C. The direction of the movement is then reversed and the cantilever moves away from the sample. At point D the cantilever feels the maximum adhesive force, upon a further increase of the distance between the cantilever and the sample the force on the tip will decrease until the cantilever returns to its rest position when the sample is once again far away[40].

Basically any point along this force curve can be used to create an image with an AFM. One fundamental distinction is whether the cantilever tip actually touches the surface of the sample or not, called *contact* and *noncontact mode*, respectively. The second fundamental distinction is whether the cantilever is oscillated or not, called *AC* and *DC mode*, respectively. In the case of *AC contact mode* the cantilever is set into oscillation close to its resonant frequency. Upon approaching the sample surface with the cantilever the oscillation amplitude is damped as soon as the tip interacts with the surface. A certain amount of damping is allowed and used as a feedback signal while the x- and y-piezos are used to track the tip along the surface. The z-piezo is controlled by this feedback loop and keeps the tip at a constant distance to the surface by keeping the amount of damping of the cantilever oscillation the same. The movements of the piezos are recorded and can be used to create an image of the sample's surface[40].

There are several methods that can be used to monitor the amplitude of the cantilever oscillation. In the case of the *optical deflection technique* the light from a laser is reflected off the back of the cantilever onto a *position sensitive detector.* This is a photodiode which emits a signal proportional to the position of a point of light upon its front surface (see Figure 3.2)[40].

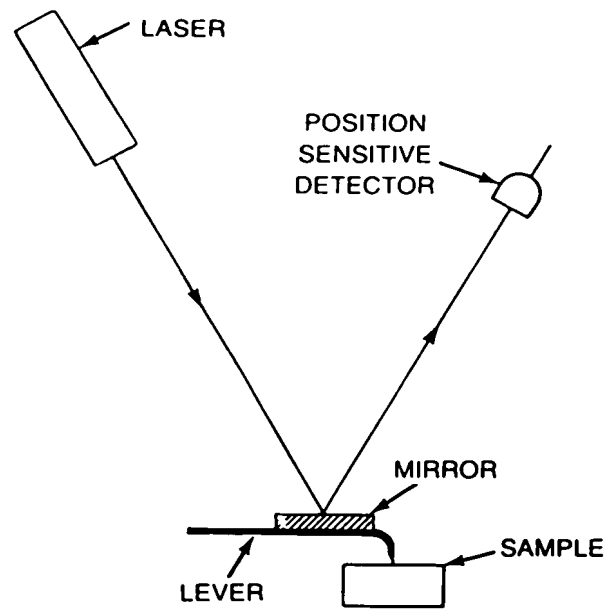

Fig. 3.2: Schematic diagram of optical deflection technique (taken from Ref. 40).

### 3.3. Electrostatic Force Microscopy:

The technique of Electrostatic Force Microscopy (EFM) is used to measure and image electric fields in a sample[40,43]. It requires the use of a metal-coated cantilever and the combination of an *AC contact mode* which determines the morphology of the sample's surface and an *AC noncontact mode* which determines the electric field, i.e. each section of the surface is traced twice. The first scan trace uses the AC contact mode described in chapter 3.2 to record the topography. For the second scan the tip is lifted to a predefined height and forced to trace the topographic structure, which decouples the topography from long-range electrostatic forces (see Figure 3.3)[47,48,49,50,51]. There are two diffe-rent principles that can be used to image the electric field in that second scan trace, the technique of *Electric Field Gradient Imaging* (EFGI) and that of *Surface Potential Imaging* (SPI).

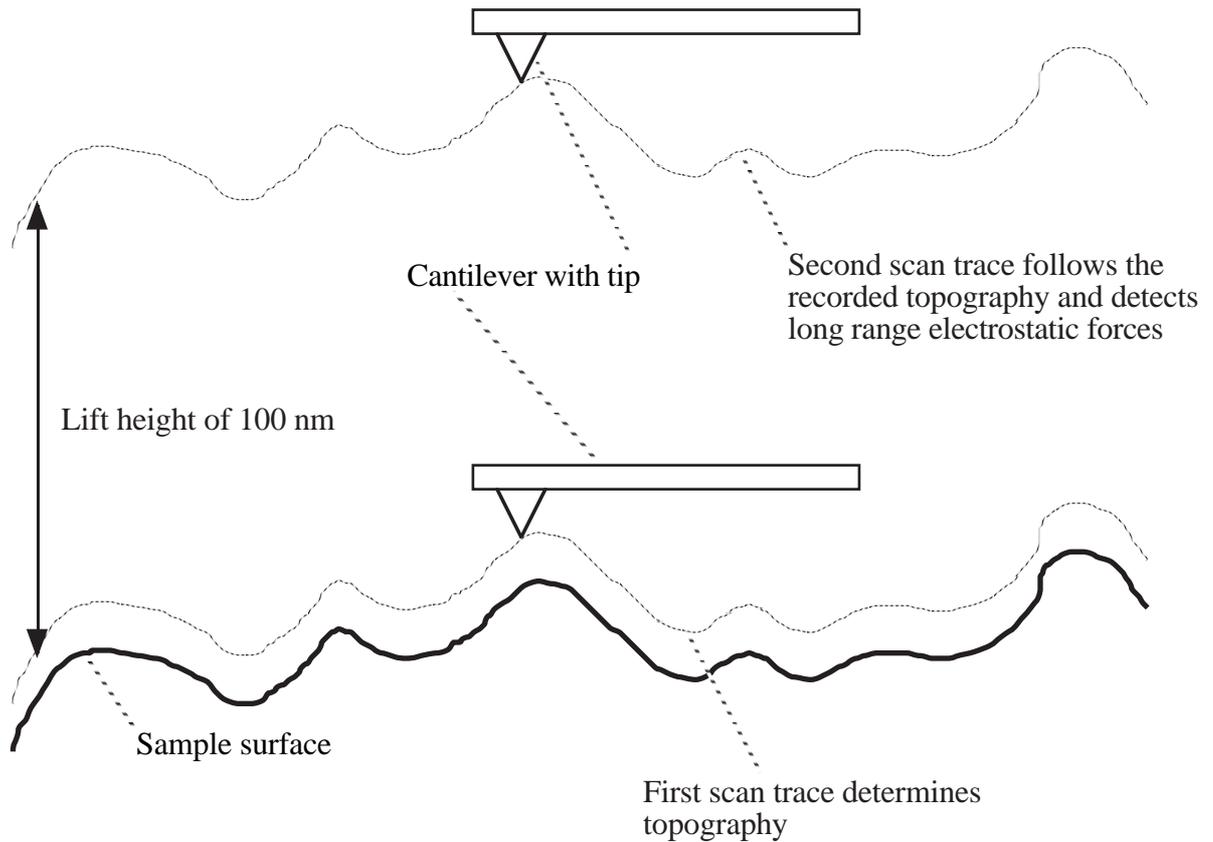

Fig. 3.3: Schematic diagram of the combined AC contact mode that acquires the topography and the reverse topography trace that detects long-range forces due to the electric field.

In the technique of EFGI the cantilever is oscillated near its resonant frequency, just as in the AC contact mode described in Chapter 3.2. As the tip follows the recorded topography, variations in the gradient of the force that the electric field exerts on the metal-coated cantilever will alter its *effective spring constant*, changing its resonant frequency. This can be detected as a change in the *amplitude*, the *frequency*, or the *phase* of the oscillation. It is important to note that images acquired in this way will not represent the electric field itself, but rather the *gradient of the field*[47,48,49].

In the technique of SPI the piezo that normally vibrates the cantilever is turned off on the second scan trace. Instead, to measure the surface potential, an oscillating voltage, $V_{AC}$, is applied directly to the cantilever tip[51]:

$$V_{AC} = V_1 \times \sin(\omega \times t)$$

(3.1)

This creates an oscillating electrostatic force at the frequency $\omega$ on the cantilever. In general, the electrostatic force on a conducting tip held close to a conducting surface is given by[50,51]:

$$F = \frac{V^2}{2} \frac{\partial C}{\partial z}$$

(3.2)

In Equation (3.2), V is the voltage difference between the tip and the specimen, C the capacitance, and z the distance between tip and specimen. Since there is an oscillating voltage applied to the tip, the voltage difference V can be expressed as[51]:

$$V = V_{DC} \times V_{AC}$$

(3.3)

In Equation (3.3), $V_{DC}$ is the DC voltage difference between the tip and the sample and $V_{AC}$ is the oscillating voltage applied to the cantilever tip. Combining Equation (3.2) and (3.3) leads to:

$$F = \frac{(V_{DC} \times V_{AC})^2}{2} \frac{\partial C}{\partial z}$$

(3.4)

Equation (3.4) shows that the force on the cantilever depends on the product of the AC drive voltage and the DC voltage difference between the tip and the sample. When the tip and sample are at the same DC voltage (which means that $V_{DC} = 0$) the cantilever will feel no oscillating force. Therefore, the local surface potential can be determined by applying a DC voltage, $V_{tip}$, to the tip and adjusting it until the oscillation amplitude of the cantilever becomes zero. At this point the tip voltage will be the same as the unknown surface potential. The value of $V_{tip}$ is recorded and used to construct a potential image of the surface[51].

### 3.4. Summary:

This chapter gave an overview of fundamental concepts important to Force Microscopy. The use of Atomic Force Microscopy to determine the topography of a sample and of Electrostatic Force Microscopy to measure an electric field in the sample were described. The theory of Surface Potential Imaging, a particular form of EFM, was explained and it was shown that SPI creates images on the basis of exact quantitative values for the local voltage of the surface.

## Chapter 4: Experimental Procedure

*One secret which I alone possessed was the hope to which I had dedicated myself; and the moon gazed on my midnight labours, while, with unrelaxed and breathless eagerness, I pursued nature to her hiding-places.*

                          Victor Frankenstein from "Frankenstein"
                          by Mary Wollstonecraft Shelley
                          Chapter 4

### **4.1. Introduction:**

This chapter presents the sample preparation and characterization that was done earlier by V. Srikant, V. Sergo, and D. Clarke. Furthermore, the principles of the technique of four point resistivity measurement after van der Pauw are explained. Finally, the exact procedure of the experiments performed on the samples is described, including the electrical measurements, the topography measurements with the AFM, and the electric field measurements using the technique of SPI.

### **4.2. Sample Preparation:**

The two samples are Al-doped ZnO thin films on $Al_2O_3$ substrates. They were prepared by V. Srikant, V. Sergo, and D.R. Clarke at the University of California, Santa Barbara[52,53]. The films were grown by pulsed laser ablation (with a laser repetition rate of 10 Hz and an energy density of 6 $J/cm^2$) of a polycrystalline ZnO pellet in a turbo-pumped vacuum system (see Figure 4.1).

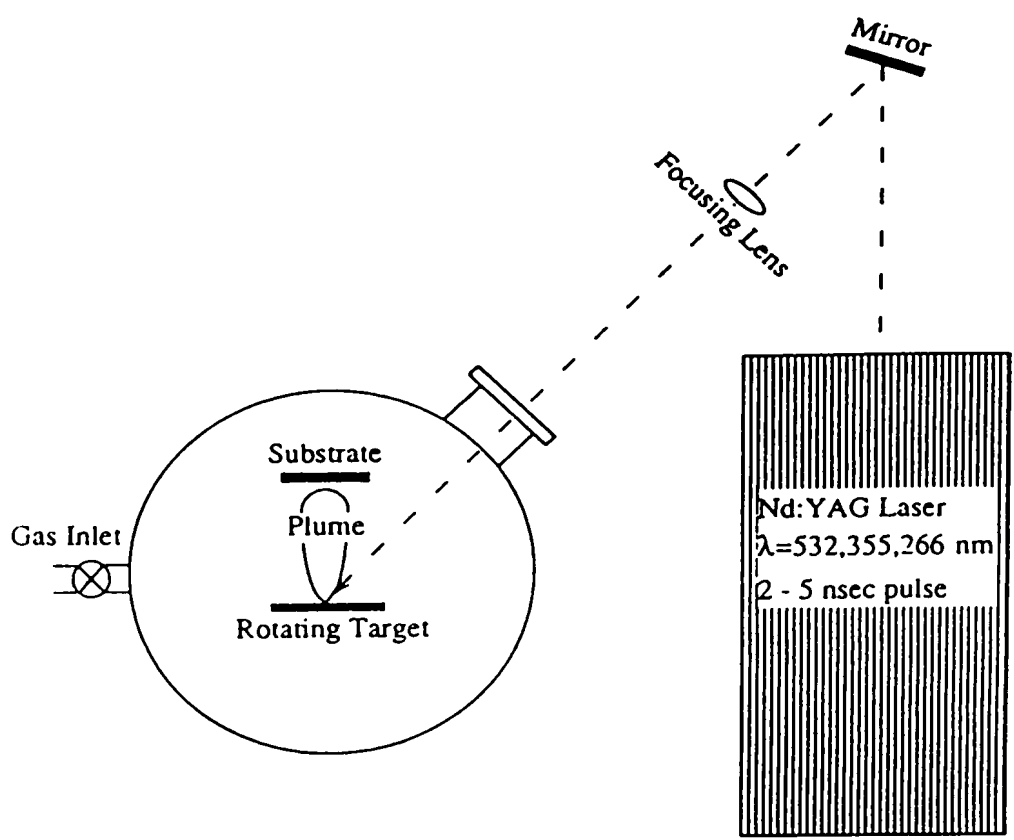

Fig. 4.1:  Schematic diagram of the laser ablation system used to prepare the ZnO thin films (taken from Ref. 52).

The targets of stoichiometric ZnO doped with 1 atom-% Al were prepared by sintering. The requisite amounts of pure ZnO and Al(NO$_3$)$_3$ were first mixed in isopropyl alcohol. The suspension was then dried in a vacuum furnace at 110 °C overnight and the resulting powder was cold-pressed at 4000 psi to form a pellet which was subsequently sintered to a density of 93 % at 1300 °C for 5 h. Each ZnO target was mounted on a multiple target holder and rotated to minimize inhomogeneous ablation. The Al$_2$O$_3$ substrates used were corundum single crystals displaying the A ($\bar{1}2\bar{1}0$) plane. Prior to growth all of the substrates were chemically cleaned and then annealed in flowing oxygen at 1400 °C for 4 h. Each substrate was then clamped onto a heater block at a distance of 5.5 cm from the target, the temperature of the substrate was monitored with a thermocouple embedded in the substrate block. Sample 1 was grown at 600 °C and Sample 2 was grown at 750 °C. The partial pressure of oxygen during the growth was dynamically maintained at 0.01 to 10 mtorr by balancing the rate of in-flow of oxygen. The growth rate of the films was approximately 2 nm/s. After the growth the substrates were cooled down to room temperature at a rate of 10 °C/min in the same partial pressure of oxygen in which the films were grown. For several electrical measurements that were later on performed on the samples, the thin films were then equipped with electrical contacts (see Figure 4.2)[52].

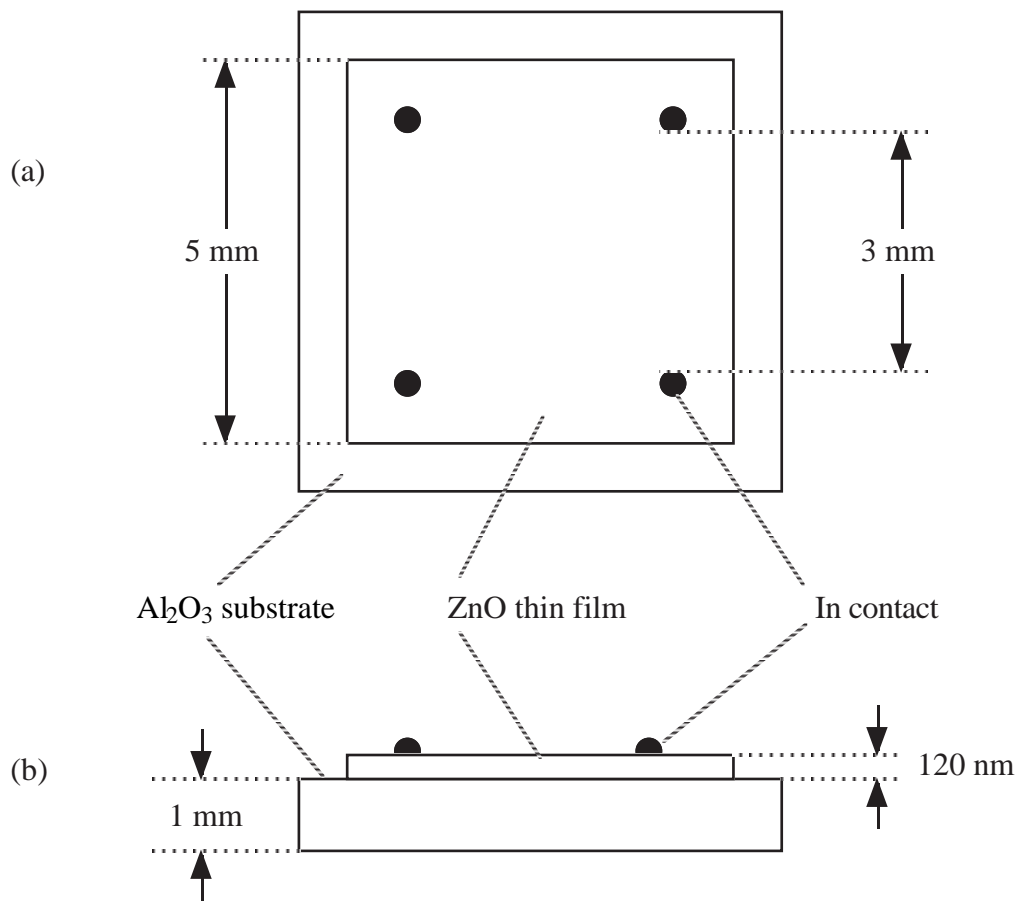

Fig. 4.2:   Schematic diagram of the samples,
(a) top view, (b) side view.

This is generally done by depositing pads of metal onto the surface of an oxide and annealing the sample. During the annealing some of the metal diffuses into the surface of the oxide and an intimate contact of the two materials is produced[54]. In this particular case pads of indium (In) were used, which is known to form good ohmic and low-resistance contacts with ZnO[55]. The two samples were prepared in the form of 5x5 mm squares with four In contacts roughly 3 mm apart from each other (see Figure 4.2)[53].

**<u>4.3. Sample Characterization:</u>**

Several experiments were already performed on the samples before this study, which have been discussed by V. Srikant, V. Sergo, and D.R. Clarke[52,53]. The thickness of the two ZnO films is approximately 120 nm as measured by ellipsometry. The crystallography of the samples was evaluated using on- and off-axis X-ray diffractometry and it was shown that the ZnO grew epitaxially, with its c-axis perpendicular to the substrate. The full width half maximum of the X-ray diffraction peaks of Sample 1 was 1.65°, for Sample 2 it was 0.57°. This indicates that the ZnO films are either single crystals with a mosaic structure or polycrystalline material with oriented grains separated by low-angle ($\approx 1°$) grain boundaries. The expected grain size of the films is 100 to 500 nm[52].

The carrier concentration and electron mobility of the films were deduced from electrical measurements performed at room temperature using the van der Pauw four point geometry. The carrier concentration for Sample 1 is $4.36 \times 10^{19}$ cm$^{-3}$, its electron mobility is 24.7 cm$^2$/Vs; in the case of Sample 2 the carrier concentration is $0.594 \times 10^{19}$ cm$^{-3}$ and the electron mobility is 10.4 cm$^2$/Vs. Sample 1 therefore exhibits an electrical behavior resembling a ZnO single crystal, while the electrical behavior of Sample 2 is rather typical of a polycrystalline ZnO material. This difference in behavior is attributed to the formation of potential barriers at the low-angle grain boundaries in the epitaxial ZnO films: the height, $\phi_B$, of the barriers is assumed to be much larger in Sample 2 than in Sample 1 (see Figure 4.3). The trap density, $N_s$, at the grain boundaries is deduced to be approximately $7 \times 10^{12}$ cm$^{-2}$ (see Figure 4.3)[53].

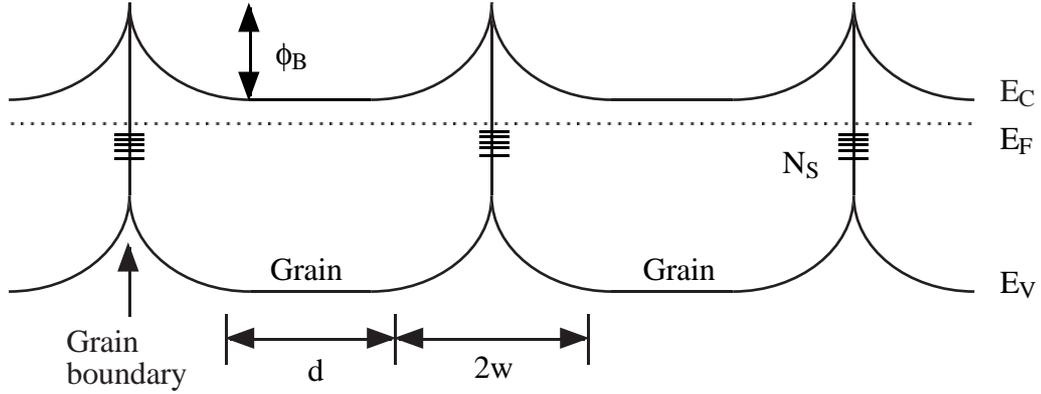

Fig. 4.3: Schematic band structure diagram for the polycrystalline ZnO films (after Ref. 53).

### 4.4. Resistivity Measurement Using the Four Point Technique:

According to L.J. van der Pauw, the specific resistivity of a flat sample can be determined with a four point geometry by measuring two resistance values and the thickness of the sample[56]. To do this, copper wires were attached to the four In contacts in the corners of the samples by using silver print. This gives four electrical contacts, A, B, C, and D (see Figure 4.4). A current, $I_{AB}$, is imposed upon the points A and B by using a power supply and measured with a multimeter set as an ammeter. At the same time, the potential drop, $V_{CD}$, that is generated across the points C and D by the flowing current through A and B is measured with a multimeter set as a voltmeter (see Figure 4.5). The value of $R_{AB-CD}$ is defined as the quotient of $V_{CD}$ and $I_{AB}$[56]:

$$R_{AB-CD} = \frac{V_{CD}}{I_{AB}}$$

(4.1)

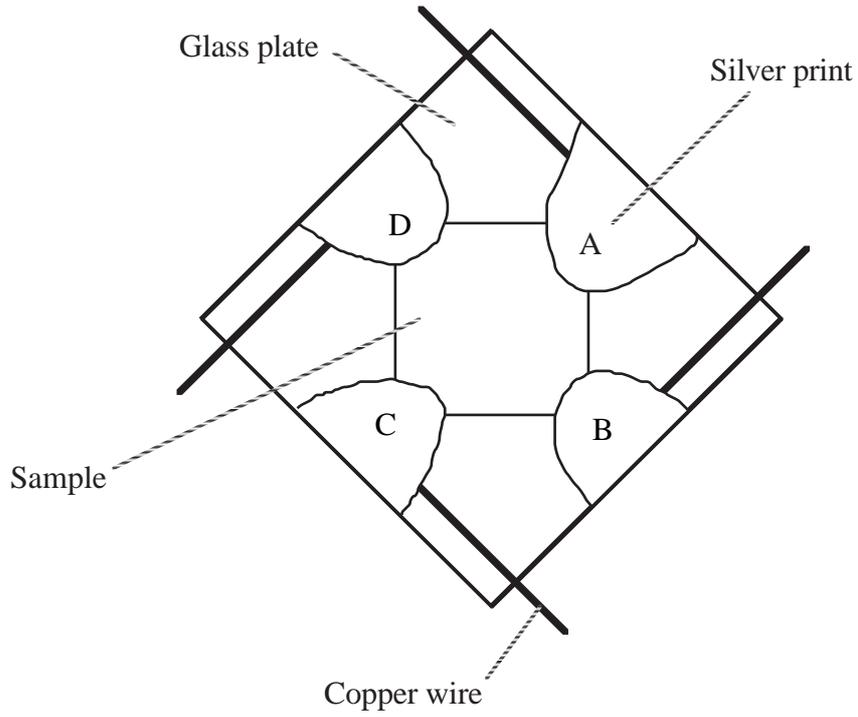

Fig. 4.4: Schematic set-up of the electrical contacts.

The resistance $R_{BC-DA}$ is measured and defined in a similar manner[56]:

$$R_{BC-DA} = \frac{V_{DA}}{I_{BC}}$$

(4.2)

Since the thickness, d, of the samples is already known, the measurement of $R_{AB-CD}$ and $R_{BC-DA}$ makes it possible to calculate the specific resistivity, $\rho$, of the ZnO thin films according to[56]:

$$\rho = \frac{\pi}{2 \times \ln 2} \times d \times (R_{AB-CD} + R_{BC-DA}) \times f$$

(4.3)

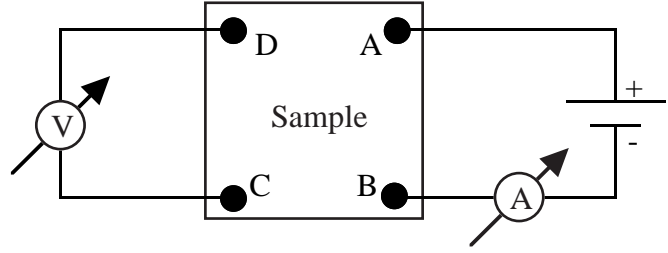

Fig. 4.5:   Measurement of $R_{AB\text{-}CD}$.

In Equation (4.3), f is a function that only depends on the ratio of $R_{AB\text{-}CD}$ to $R_{BC\text{-}DA}$ and is described by[56]:

$$\frac{R_{AB\text{-}CD} - R_{BC-DA}}{R_{AB-CD} + R_{BC-DA}} = f \times \text{arccosh}\left(\frac{e^{(\ln 2/f)}}{2}\right)$$

(4.4)

If the values of $R_{AB\text{-}CD}$ and $R_{BC\text{-}DA}$ are not too different from one another, f can be approximated with the following equation[56]:

$$f \approx 1 - \left(\frac{R_{AB\text{-}CD} - R_{BC\text{-}DA}}{R_{AB\text{-}CD} + R_{BC\text{-}DA}}\right)^2 \times \frac{\ln 2}{2} - \left(\frac{R_{AB\text{-}CD} - R_{BC\text{-}DA}}{R_{AB\text{-}CD} + R_{BC\text{-}DA}}\right)^4 \times \left(\frac{(\ln 2)^2}{4} - \frac{(\ln 2)^3}{12}\right)$$

(4.5)

### **4.5. Measurement of I-V Curves:**

I-V curves of the samples were measured with a standard two point geometry, using two of the four contacts. A voltage, V, is imposed between the points by using a power supply and measured with a multimeter set as a voltmeter. At the same time the current flowing through the points is measured with a multimeter set as an ammeter (see Figure 4.6).

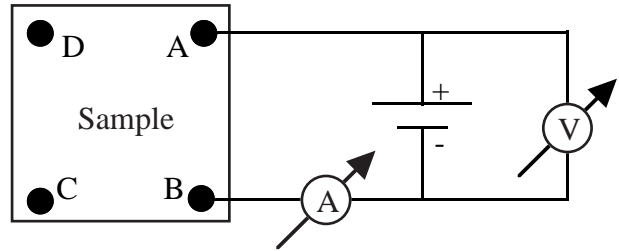

Fig. 4.6:   Measurement of the I-V curves.

## 4.6. Topography Measurement Using Atomic Force Microscopy:

Topography measurements were performed with a Digital Instruments (DI) Nano Scope Dimension 3000 Scanning Probe Microscope (SPM), using silicon cantilevers with a length of 124 µm and resonant frequencies between 320 and 440 kHz. The mode used is an AC contact mode that is called "Tapping Mode" by the manufacturer (the principle of which is described in Chapter 3.2).

## 4.7. Measurement of the Electric Field Using Surface Potential Imaging:

The combination of the DI Dimension 3000 SPM with a DI Extender Electronics Module enables the SPM to perform the technique of Surface Potential Imaging (see Figure 4.7). In this case gold-coated silicon cantilevers with a length of 225 µm and resonant frequencies between 59 and 104 kHz were used. Each scan line was traced twice, the first trace determined the topography, using the "Tapping Mode", and the second trace determined the surface potential, using a constant sample-tip separation mode which is called "Lift Mode" by the manufacturer (the principle of which is described in Chapter 3.3). The lift height was chosen to be 100 nm for all images that were acquired.

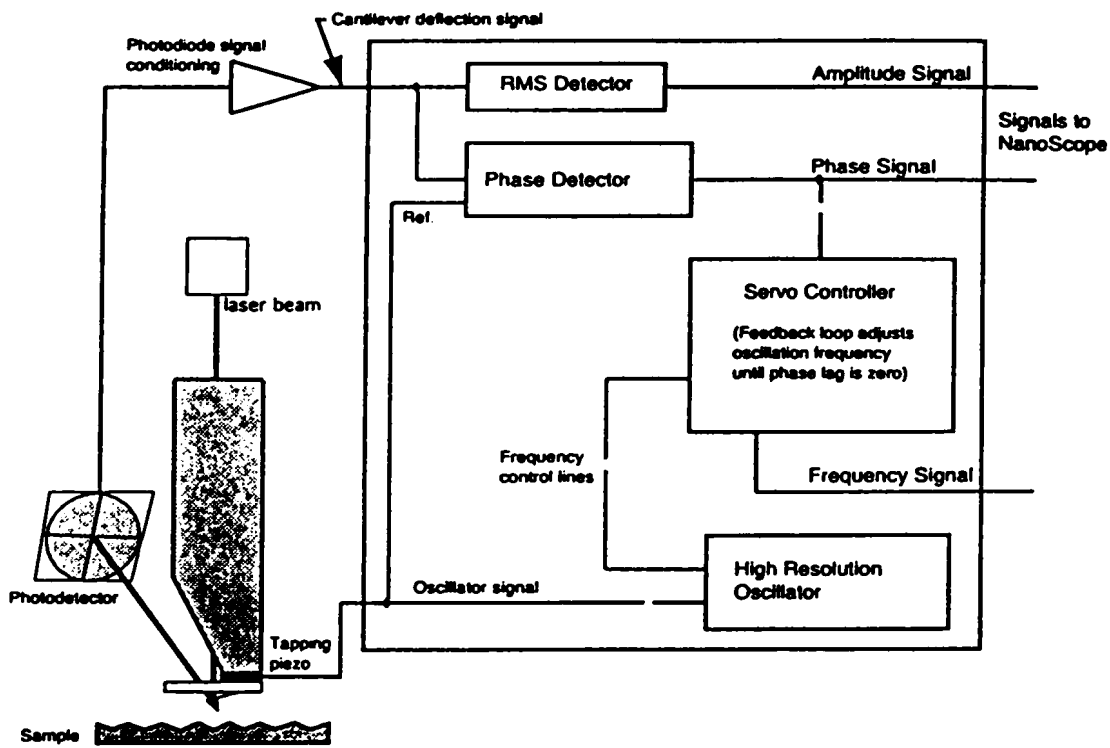

Fig. 4.7: Simplified block diagram of the surface potential measurements with the Extender Electronics Module. This module is configured in such a way that the phase detection circuitry acts as a lock-in amplifier.

An external voltage was applied to the samples with a power supply via two neighbors of the four contacts, and was measured with a multimeter set up as a voltmeter. In order to keep electronic noise to a minimum, the ground terminal of the power supply was connected to the ground pin of the SPM (see Figure 4.8). The surface potential images were acquired close to one edge of the samples, in the middle between the two contacts that were used to apply the voltage to the sample (see Figure 4.9).

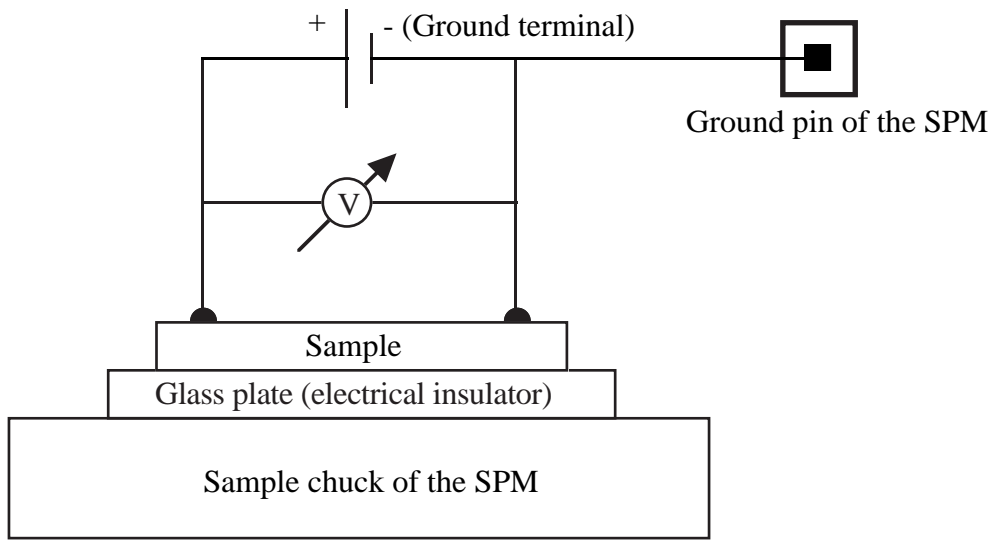

Fig. 4.8: Schematic set-up of the SPI measurements.

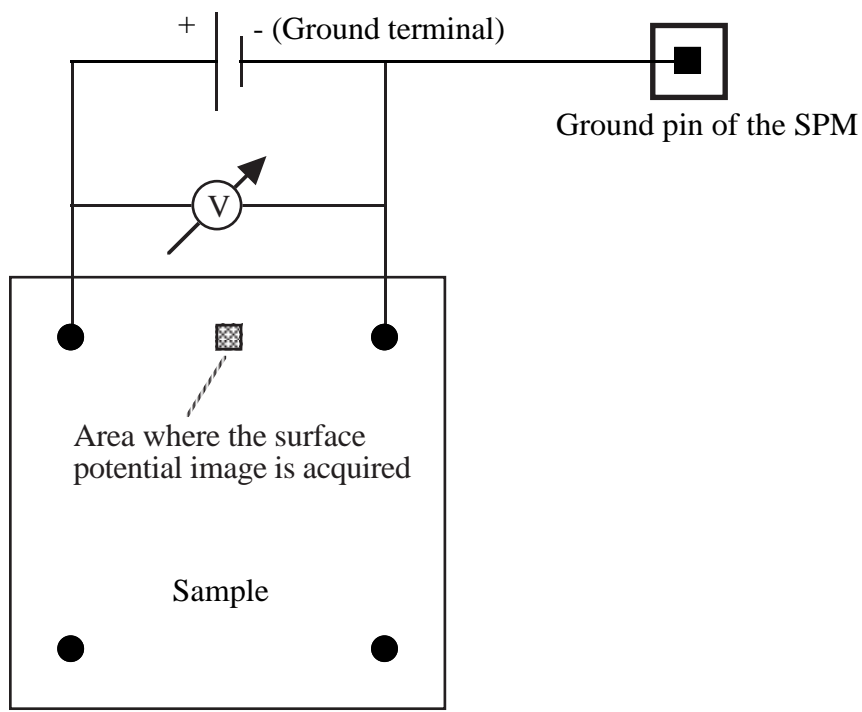

Fig. 4.9: Schematic diagram of the location used to perform the SPI measurements.

**4.8. Summary:**

This chapter presented the preparation of the ZnO thin films and their characterization, that was done before this study by V. Srikant et al. It was shown that the structure of the two samples is assumed to be polycrystalline with low angle grain boundaries and a grain size between 100 and 500 nm. Furthermore, it was hypothesized by V. Srikant et al. that there are potential barriers at the grain boundaries, which are supposedly much higher in Sample 2 than in Sample 1.

This chapter also briefly explained the principle of the technique of four point resistivity measurement after van der Pauw and presented the exact procedure of the experiments performed on the samples in this study. These are electrical measurements, topography measurements with the AFM, and electric field measurements using the technique of SPI.

## Chapter 5: Results

*[A] scientist must also be absolutely like a child. If he sees a thing, he must say that he sees it, whether it was what he thought he was going to see or not. See first, think later, then test. But always see first. Otherwise you will only see what you were expecting. Most scientists forget that.*

> John Watson from "So long, and thanks for all the Fish"
> by Douglas Adams
> Chapter 31

### 5.1. Introduction:

This chapter presents the results of the experiments performed on the two ZnO samples. The electrical properties of the samples are described, as are the morphology of their surfaces, and the outcome of the SPI measurements. Furthermore, the results of SPI measurements on control samples (copper and silicon) are shown, which are used to estimate the error of the technique.

### 5.2. Electrical Properties:

The resistivities, ρ, of the samples, as measured with the four point technique after van der Pauw, are $\rho^{Sam\ 1} = (2.3\pm0.1)\times10^{-3}\ \Omega\times cm$ for Sample 1 and $\rho^{Sam\ 2} = (1.02\pm0.07)\times10^{-2}\ \Omega\times cm$ for Sample 2. These values are the average values of several measurements that were done with each sample, using different applied voltages. The error is taken as the largest deviation of the individual measurements from that average value.

Current-voltage (I-V) curves for the samples, as measured with a standard two point geometry, are shown in Figures 5.1 and 5.2. It can be seen that the I-V curve of Sample 2 (Figure 5.2) is linear along the entire voltage range, whereas the I-V curve of Sample 1 is linear for low voltages and nonlinear for higher voltages (Figure 5.1). The curve trails off to higher current values, i.e. the conductivity of Sample 1 gets higher at a large voltage.

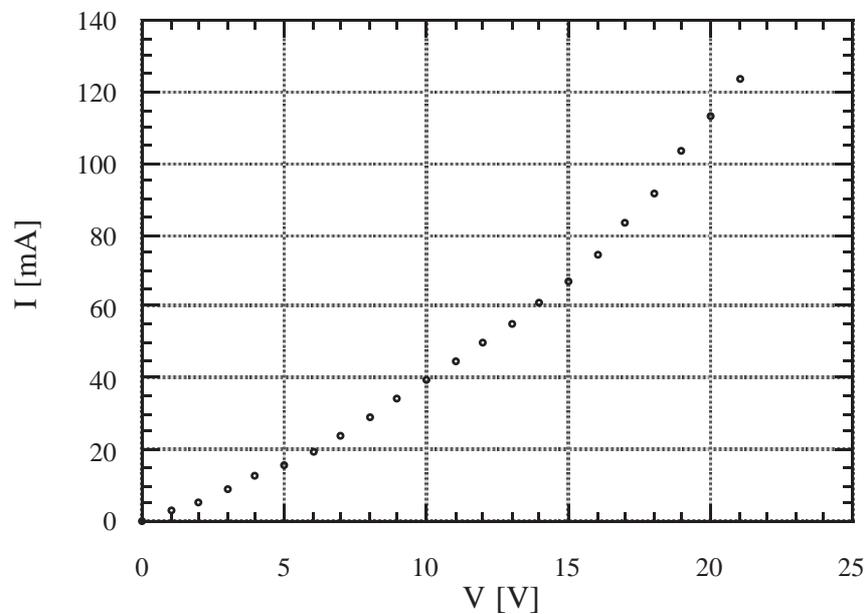

Fig. 5.1: I-V curve of Sample 1.

For both samples the measurement was stopped at the highest voltage value indicated in the figures (21 V for Sample 1 and 70 V for Sample 2) because the flowing current heated the samples up, according to *Joule's law*[57], and the effect got worse with increasing voltage. At the indicated values the temperature of the samples became so high that the organic binder in the silver print that was used to establish an electrical contact between the wires and the In contacts on the samples began to evaporate, thus changing the properties of the contacts themselves. Under these conditions the data can not be evaluated, due to the convolution of the effects taking place in the sample and in the contacts.

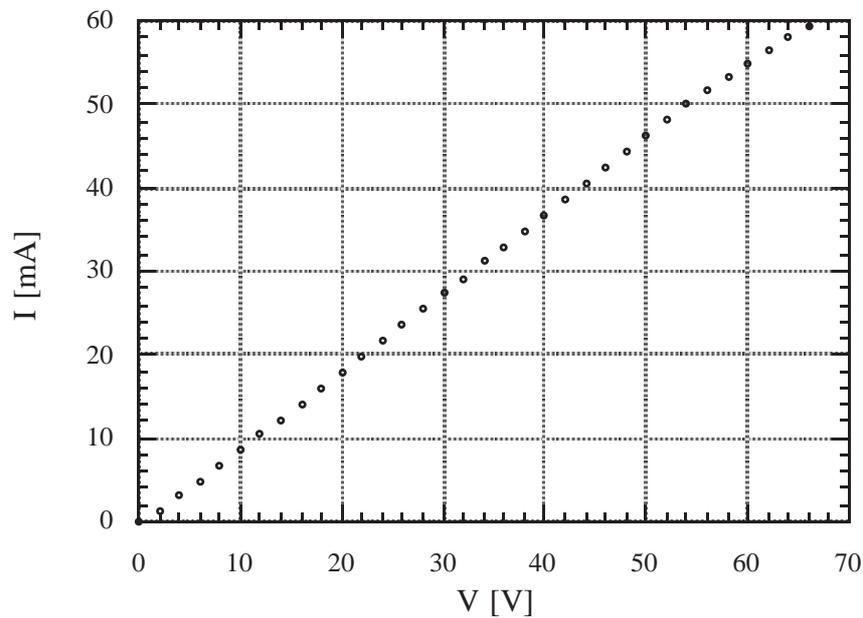

Fig. 5.2: I-V curve of Sample 2.

**5.3. Topography:**

Small scale AFM images of the two samples show a surface dominated by spherical elevations with a submicron diameter, see Figures 5.3 and 5.4. In Sample 2 the elevations are closely spaced together, forming features that look like clusters (see Figure 5.4), whereas in Sample 1 the elevations seem to be separated from one another by regions of the surface that are very flat (see Figure 5.3).

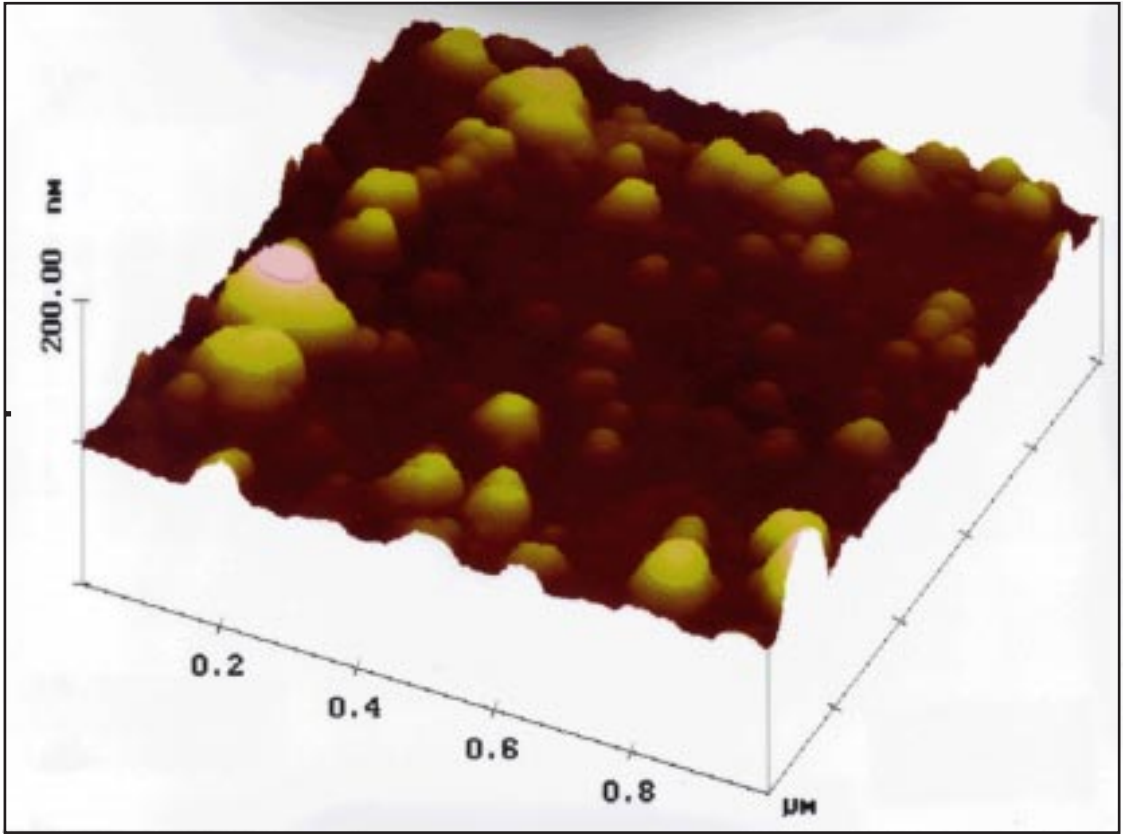

Fig. 5.3: AFM image of Sample 1, dimension of image is 1 μm X 1 μm. The z-axis (height) shows values from 0 to 200 nm.

Sections of several images were evaluated to obtain estimates of the average diameters, $d_{elev}$, and average heights, $h_{elev}$, of the elevations. The elevations in Sample 1 have an average diameter of $d_{elev}^{Sam\ 1} = (130\pm40)$ **nm** and an average height of $h_{elev}^{Sam\ 1} = (32\pm13)$ **nm**. For Sample 2 the average diameter is $d_{elev}^{Sam\ 2} = (150\pm30)$ **nm** and the average height is $h_{elev}^{Sam\ 2} = (39\pm19)$ **nm**. The errors are taken as the largest deviation of individual measurements from the average values.

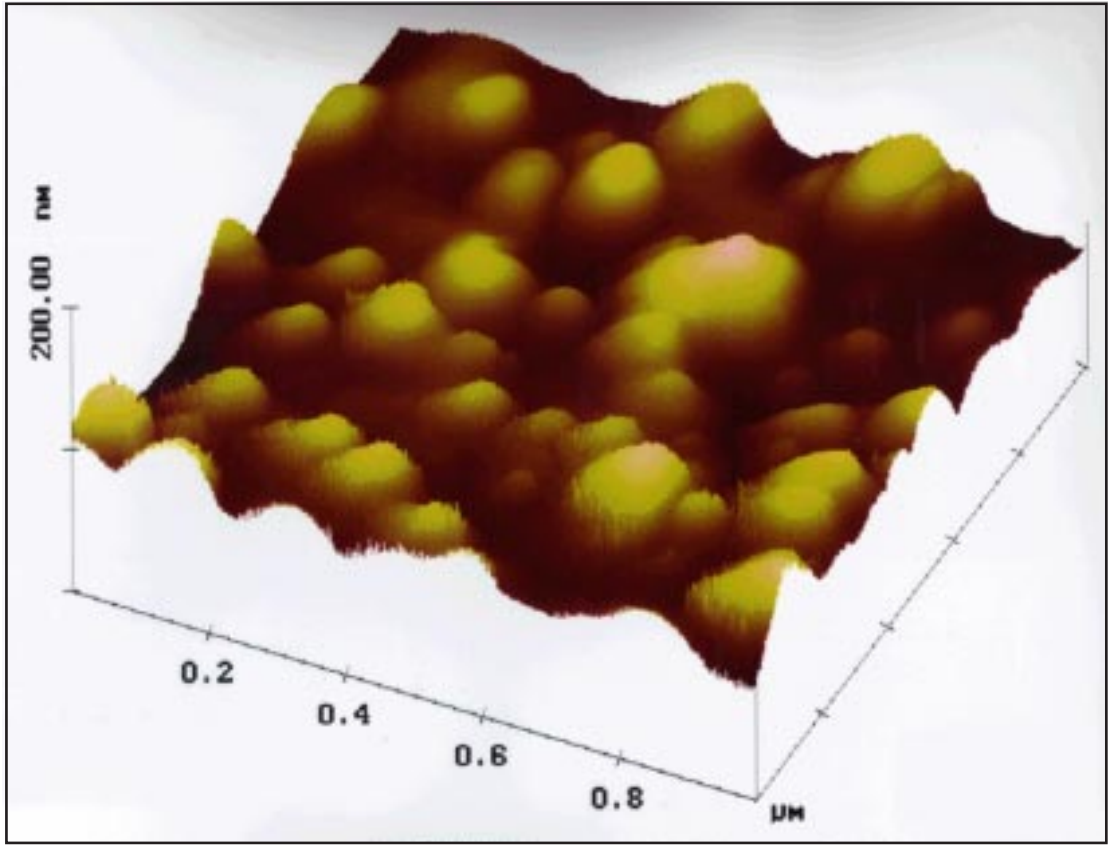

Fig. 5.4: AFM image of Sample 2, dimension of image is 1 μm X 1 μm. The z-axis (height) shows values from 0 to 200 nm.

Larger scale images show two more features, large, high protrusions and channels, that are scattered across the surface, see Figures 5.5 and 5.6. Sections of several images were evaluated to get estimates of the average diameters, $d_{pro}$, and average heights, $h_{pro}$, of the protrusions. The protrusions on Sample 1 have an average diameter of **$d_{pro}^{Sam\ 1} = (1700 \pm 1100)$ nm** and an average height of **$h_{pro}^{Sam\ 1} = (280 \pm 110)$ nm**. For Sample 2 the average diameter is **$d_{pro}^{Sam\ 2} = (1300 \pm 700)$ nm** and the average height is **$h_{pro}^{Sam\ 2} = (170 \pm 130)$ [nm]**. The errors are taken as the largest deviations of individual measurements from the average values.

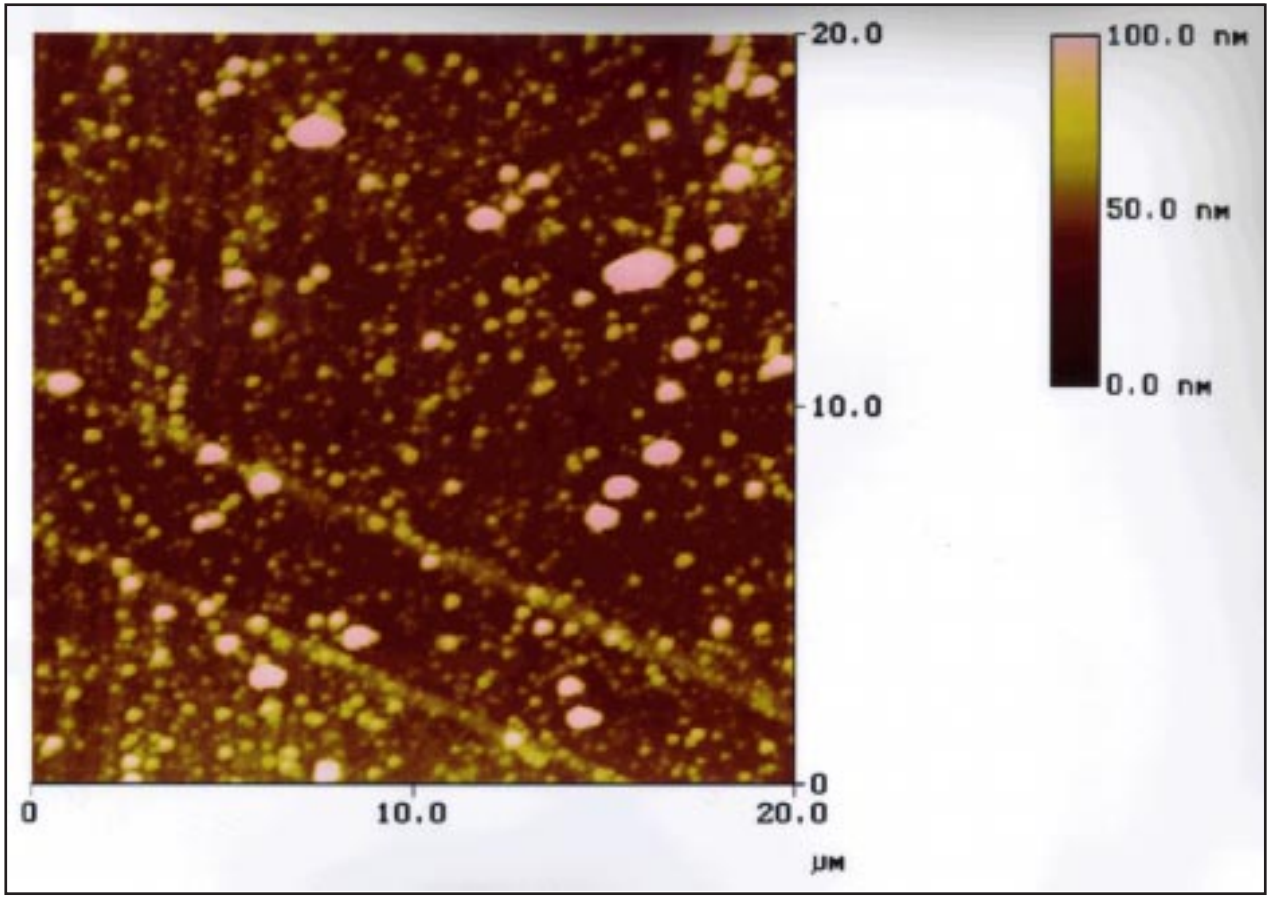

Fig. 5.5: AFM image of Sample 1, dimension of image is 20 μm ✗ 20 μm. The z-axis (height) shows values from 0 nm (designated by a bright yellow color) to 100 nm (designated by a dark brown color).

The protrusions were first interpreted as dust particles on the surfaces of the samples. For that reason the samples were cleaned for five minutes in each acetone and methanol using an ultrasonic cleaner, but the protrusions were still present afterwards. Therefore, they are probably real features in the surface structures. The channels are most probably scratches in the surface of the $Al_2O_3$ substrate. Since the ZnO film is very thin, approximately 120 nm thick, it traced out the scratches in the substrates during the deposition, forming channels.

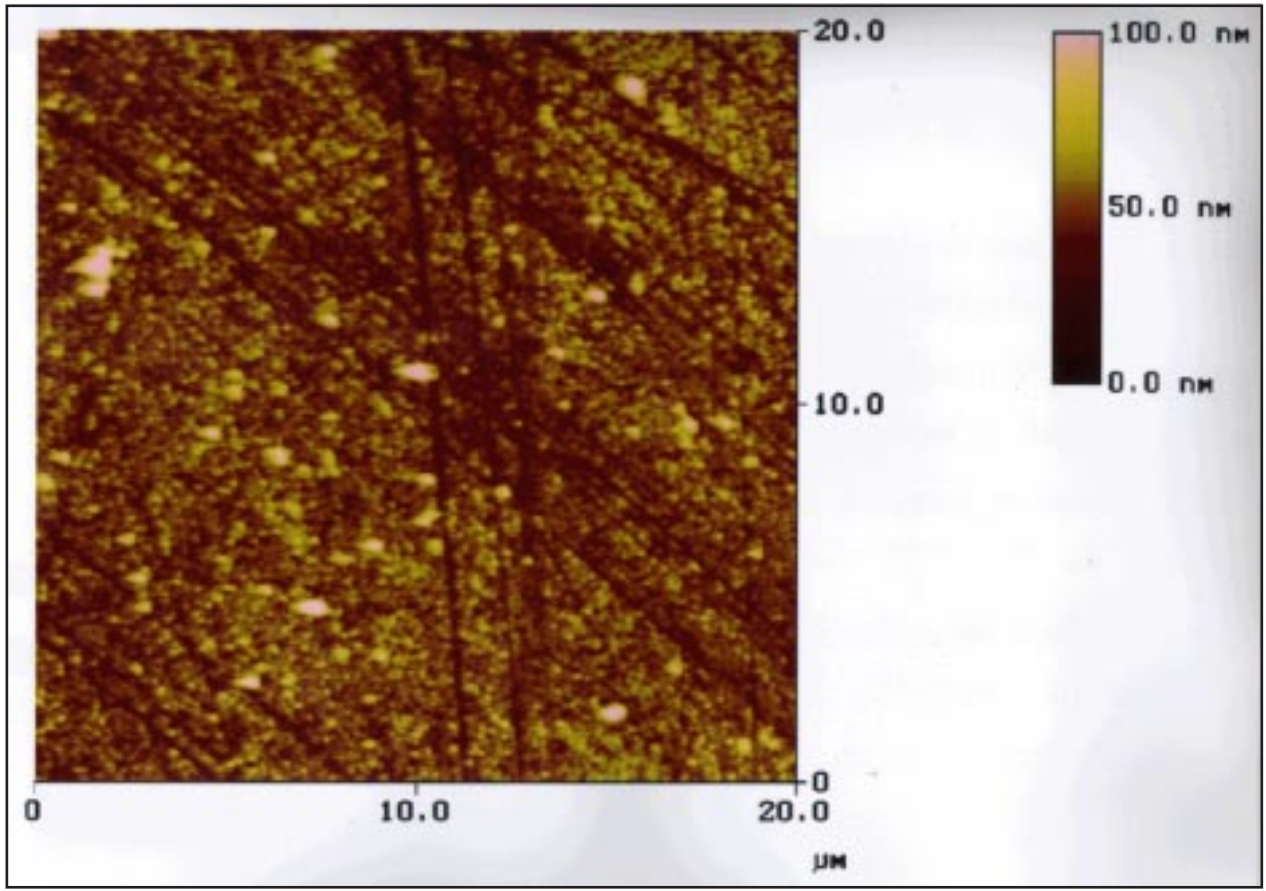

Fig. 5.6: AFM image of Sample 2, dimension of image is 20 μm X 20 μm. The z-axis (height) shows values from 0 nm (designated by a bright yellow color) to 100 nm (designated by a dark brown color).

## **5.4. Surface Potential Images Without an Applied Electric Field:**

To determine the possible contribution of topography, SPI measurements were made with two control samples while no external electric field was applied. One of the two samples was pure copper (Cu) that was polished to 50 nm with alumina grit; the other was a silicon (Si) wafer. Neither sample should show any features in the SP image as long as no electric field is applied: the Cu sample is polycrystalline, but the high conductivity of Cu does not allow the accumulation of charge at grain boundaries; the Si wafer is a single crystal and therefore has no grain boundaries. In contrary to that consideration, both samples showed features rather than just giving flat images (see Figure 5.7).

Several sections of the images were evaluated to get estimates of the maximum difference in surface potential, $\Delta V_{max}$, and compare them to the maximum difference in height, $\Delta h_{max}$, in the corresponding topography image. The maximum difference in SP for the Cu sample was $\mathbf{\Delta V_{max}^{Cu} = 0.19\,V}$, corresponding to a height difference of $\mathbf{\Delta h_{max}^{Cu} = 45\,nm}$ in the topography image. In the case of Si, the maximum difference in SP is $\mathbf{\Delta V_{max}^{Si} = 0.18\,V}$, corresponding to a height difference of $\mathbf{\Delta h_{max}^{Si} = 70\,nm}$. Figure 5.7 shows that the SP images look like an inverse of the according topography images. The topography image (left side of Figure 5.7), presents several more or less spherical elevations that appear as bright spots. The SP image (right side of Figure 5.7), shows the same features, but this time as pits, appearing as dark spots.

The first attempts at attaining SP images without an applied voltage of the two ZnO samples were not successful, because the wires were already attached to the samples but not grounded during the experiment. The wires picked up electro-magnetic impulses, basically operating as antennas[57], and therefore the images only contained oscillatory noise. This problem was solved by grounding the wires, and subsequently SP images were taken of Sample 2 (see Figure 5.8).

The SP images again look like an inverse of the topography images. The left side of Figure 5.8 shows the known morphology of Sample 2, clusterlike spherical elevations, two high protrusions, and two channels that seem to form an upside down "Y". The SP image (right side of Figure 5.8), shows the inverse of these features: pits instead of the elevations and protrusions, and ridges instead of channels.

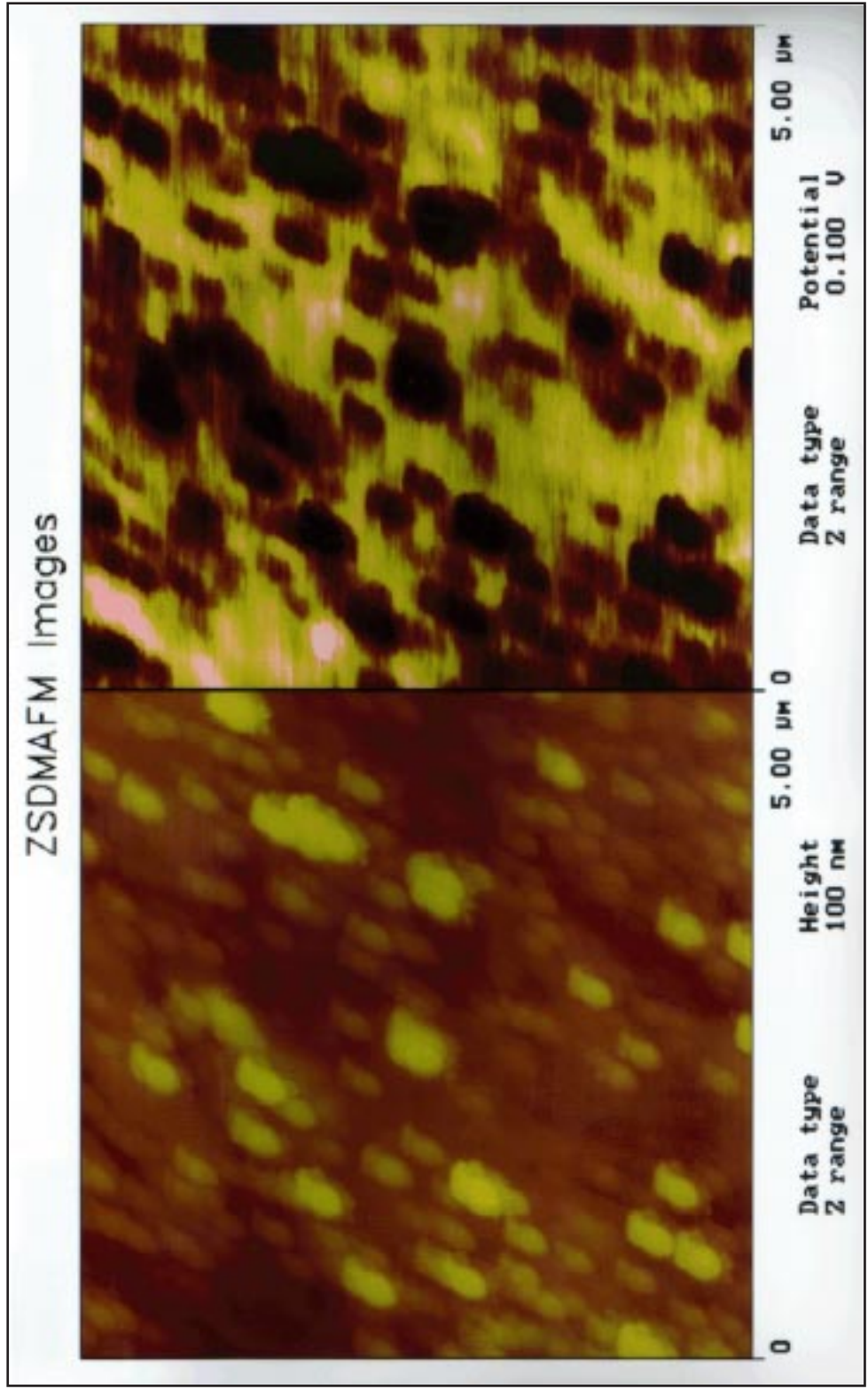

Fig. 5.7: Topography image (left side) and SP image (right side) of Cu sample, no external applied electric field.

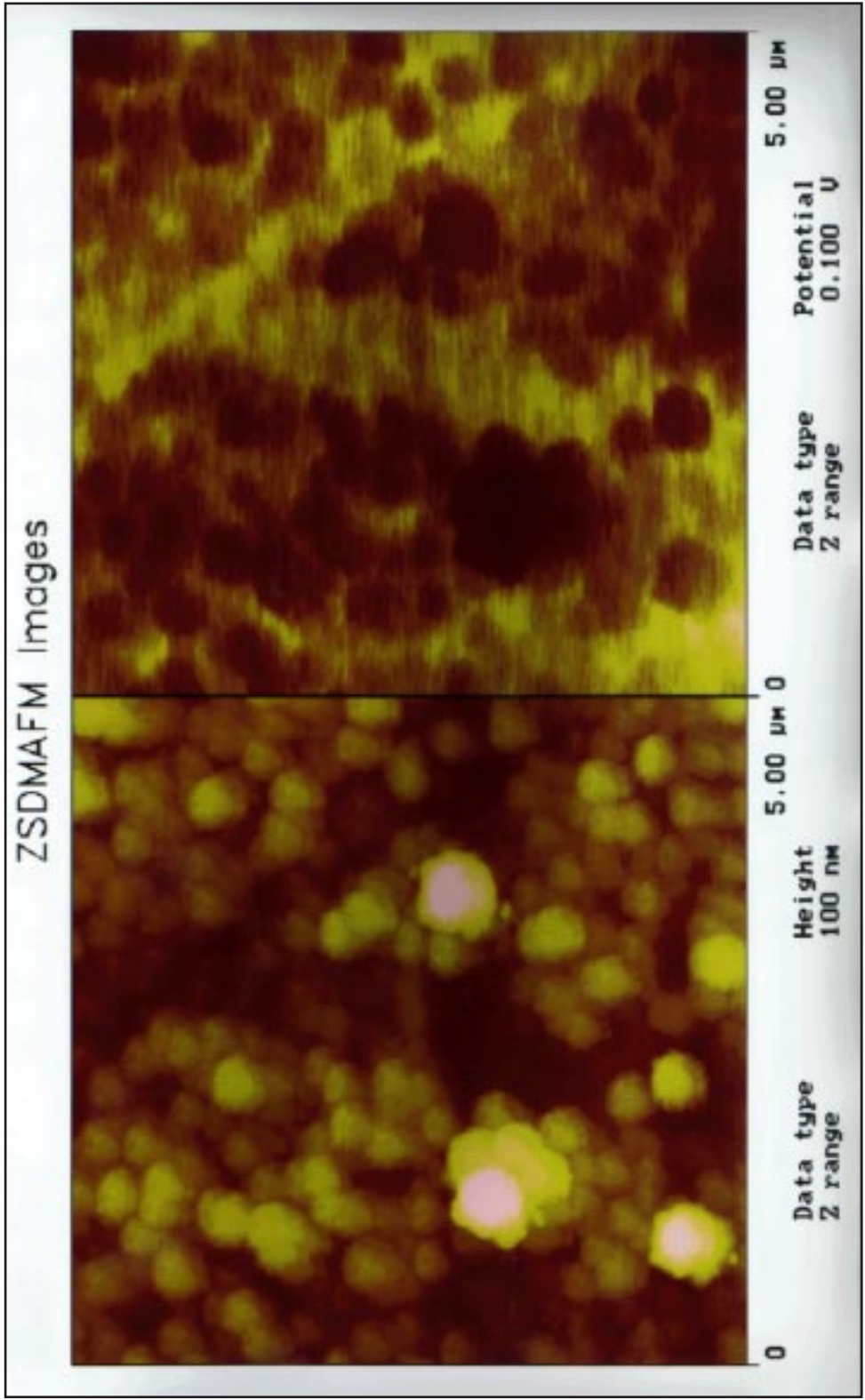

Fig. 5.8: Topography image (left side) and SP image (right side)
Sample 2, no external applied electric field.

Sections of the images revealed that the maximum difference in SP for the Sample 2 was $\Delta V_{max}^{Sam\ 2}$ = **0.066 V**, corresponding to a height difference of $\Delta h_{max}^{Sam\ 2}$ = **80 nm** in the topography image. It should be noted that the value of $\Delta V_{max}^{Sam\ 2}$ is smaller than the values of $\Delta V_{max}^{Cu}$ and $\Delta V_{max}^{Si}$, about one third of the value, even though $\Delta h_{max}^{Sam\ 2}$ is actually slightly larger than $\Delta h_{max}^{Cu}$ and $\Delta h_{max}^{Si}$. The reason for this is probably that Sample 2 was grounded, which was not the case for the Cu and the Si sample. For that reason any charges that might accumulate on the sample, e.g. static electricity, are dissipated away from it. Grounding the sample obviously improves the accuracy of the technique.

### 5.5. Surface Potential Images With an Applied Electric Field:

Since the value of $\Delta V_{max}^{Sam\ 2}$ is rather large compared to the expected voltage drop across a given length of a sample, the SP images with an applied electric field were acquired at a larger scale (80 µm). These images also show features of the topography. Just as before the SP images look like an inverse of the corresponding topographic features (see Figure 5.9). The topography image (left side of Figure 5.9), shows two clusters of features (at the lower right corner of the image) that form large elevations with diameters of roughly 20 µm. The SP image (right side of Figure 5.9), shows an inverse of that feature, the two clusters form large pits with diameters of roughly 20 µm.

In addition to that, the images that were acquired while an external electric field was applied show a slope from high voltage values (at the side closer to the positive contact) to low voltage values (at the side closer to the negative contact). This can be seen in Figure 5.9, but the effect is more obvious in the Figures 5.10 and 5.11.

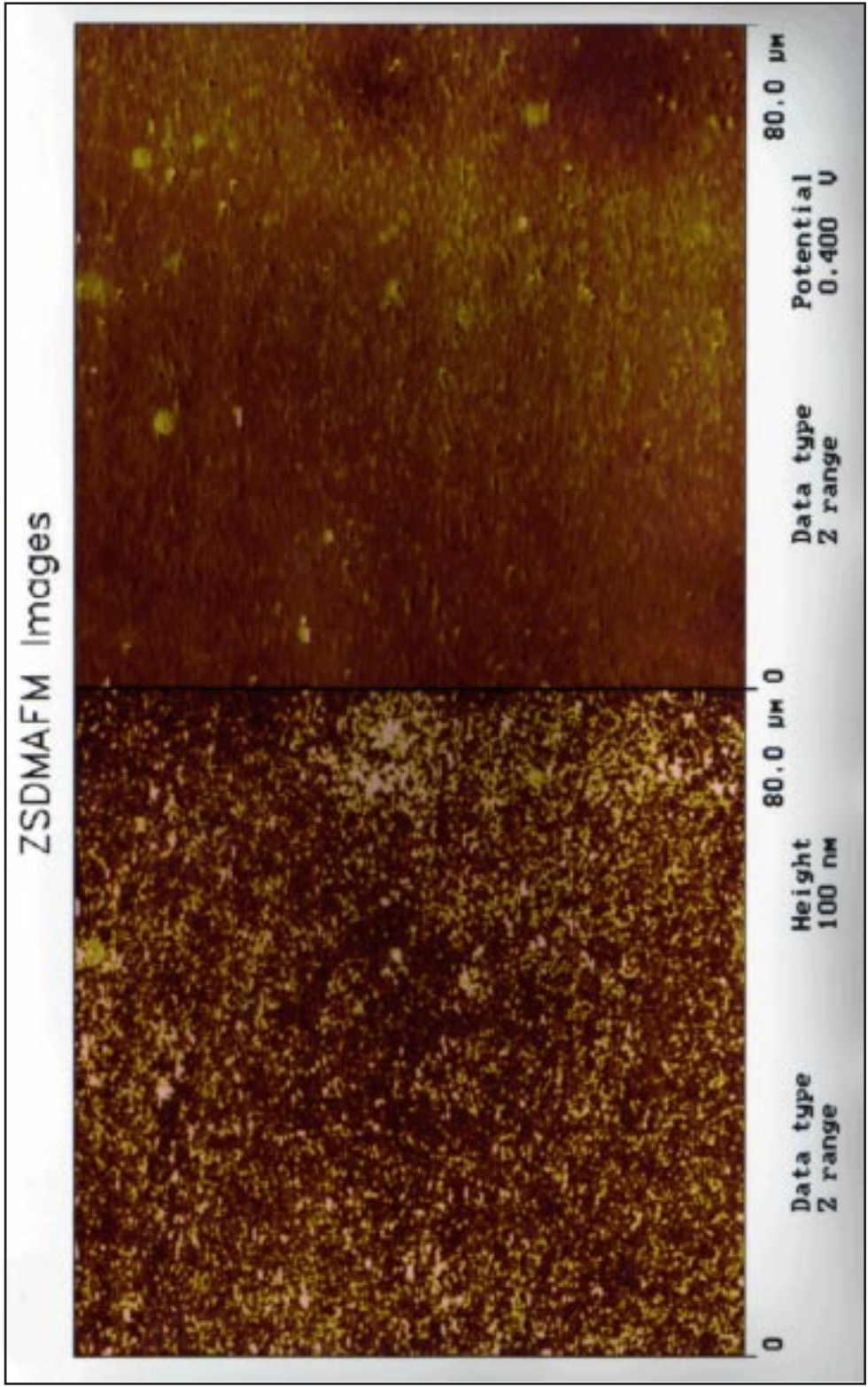

Fig. 5.9: Topography image (left side) and SP image (right side) of Sample 1, applied voltage: 3 V. Positive contact is beyond the right side, negative contact is beyond the left side.

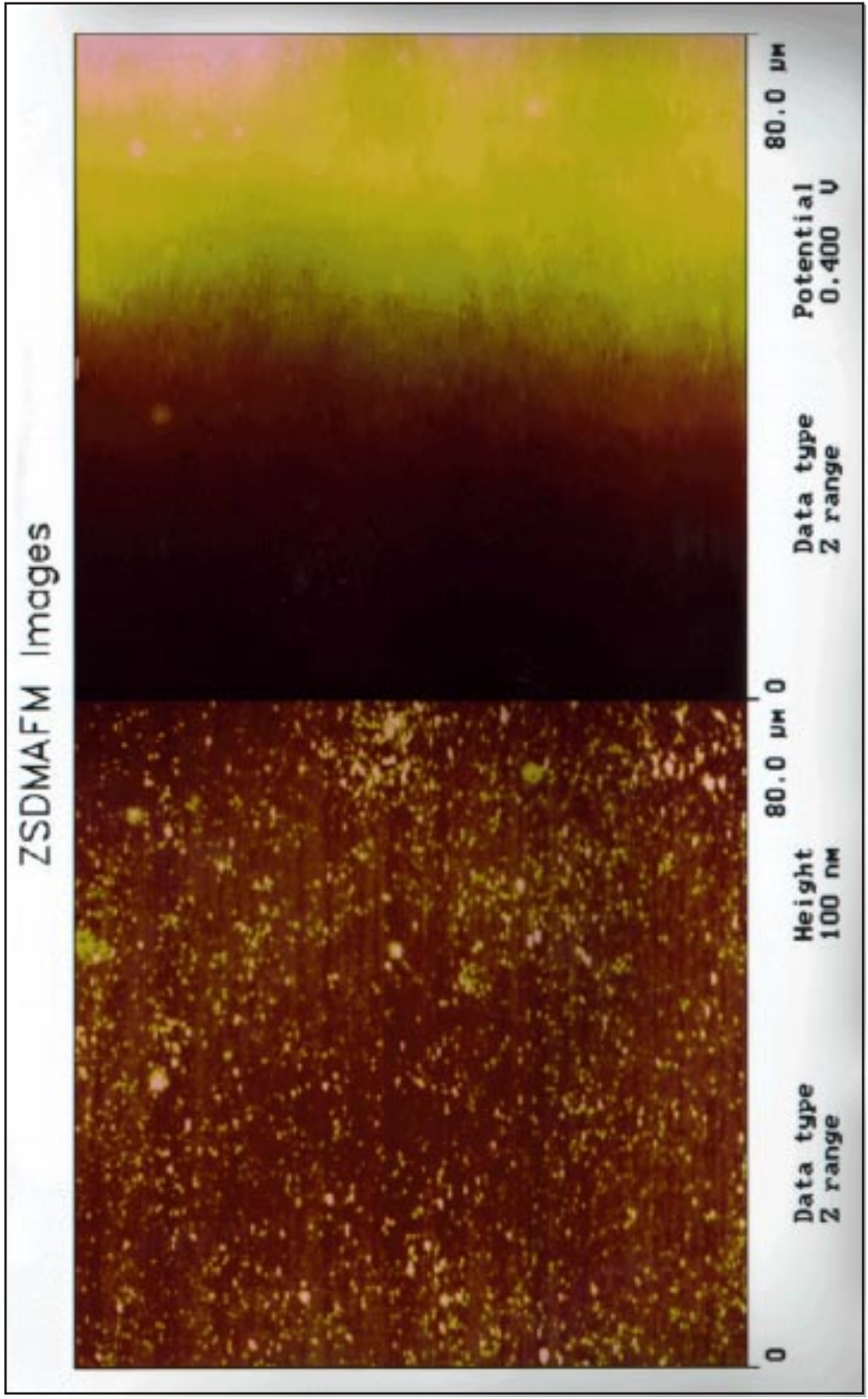

Fig. 5.10: Topography image (left side) and SP image (right side) of Sample 1, applied voltage: 15 V. Positive contact is beyond the right side, negative contact is beyond the left side.

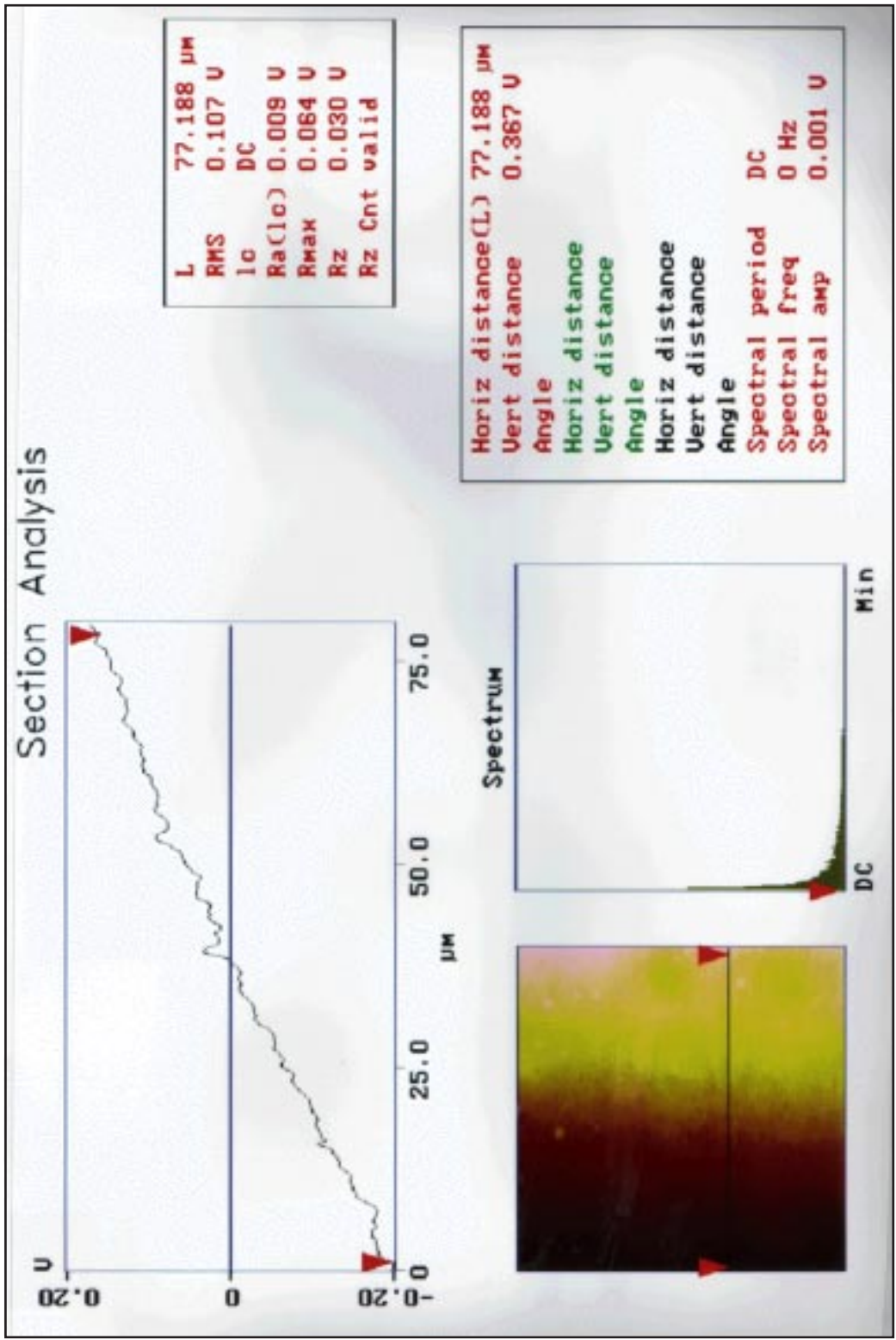

Fig. 5.11: Section of the SP image shown in Fig. 5.10.

The Figures 5.9 and 5.10 were taken at the same spot of Sample 1 and should, therefore, show the same topography. This is not absolutely true, the two images are slightly different. The distortion between a topographic image obtained with no applied voltage and one obtained with an applied voltage gets more severe with higher voltage. The images in Figure 5.12 and 5.13 were acquired at the same spot of Sample 2, one with no applied voltage, the other with an applied voltage of $V_{app} = 25$ V.

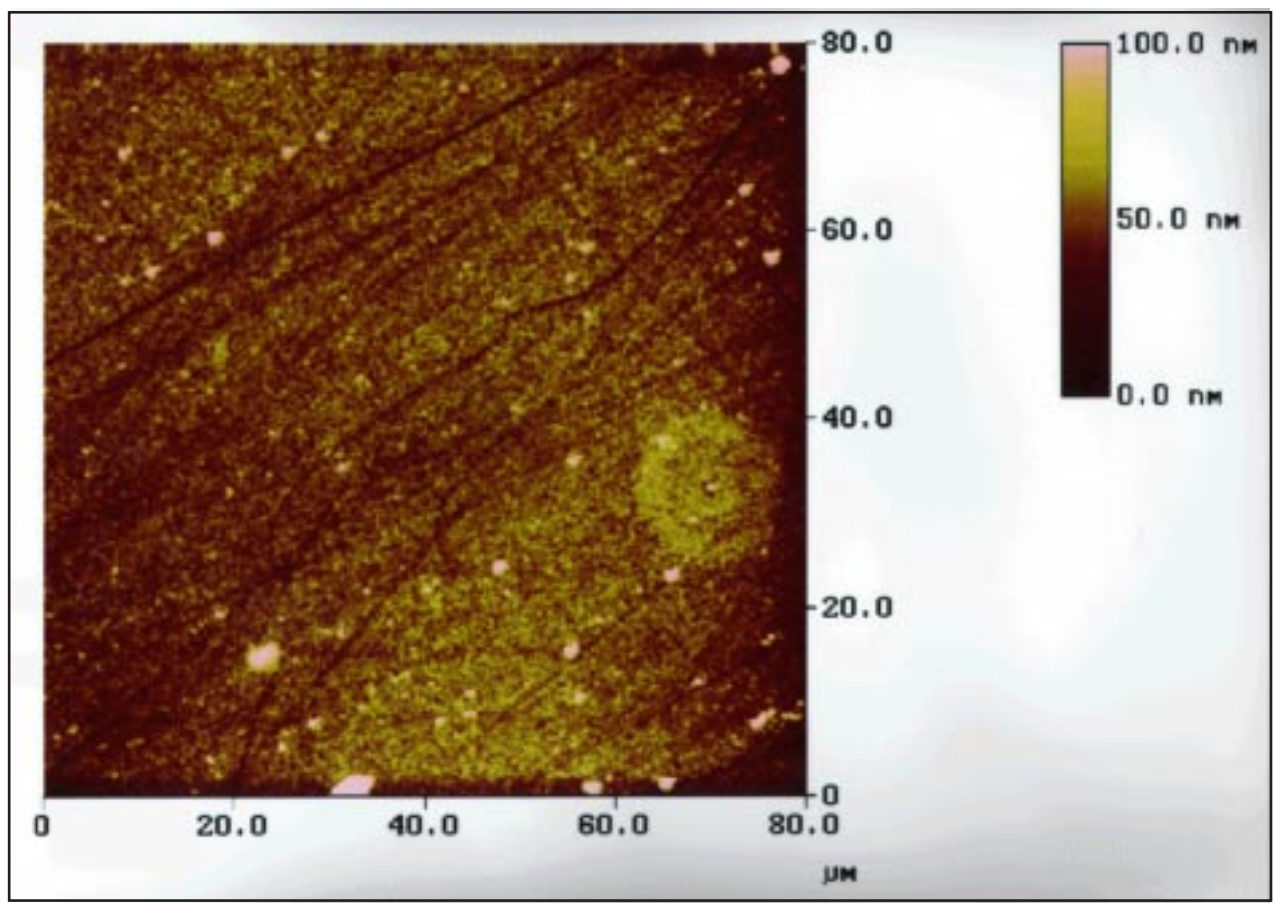

Fig. 5.12: Topography image of Sample 2, applied voltage: $V_{app} = 0$ V.

There is a noticeable difference between the two images, in particular the features of the image with no applied voltage, Figure 5.12, are rather distinct, whereas they seem blurred and smeared in the image with the high applied voltage, Figure 5.13. Since the topography data is the starting point of the SP data, the whole measurement does not provide valuable data as soon as the distortion of the topography image due to the voltage becomes too large. For this reason, quantitative data was only taken from images obtained with an applied voltage of $V_{app}$ = 15 V or less.

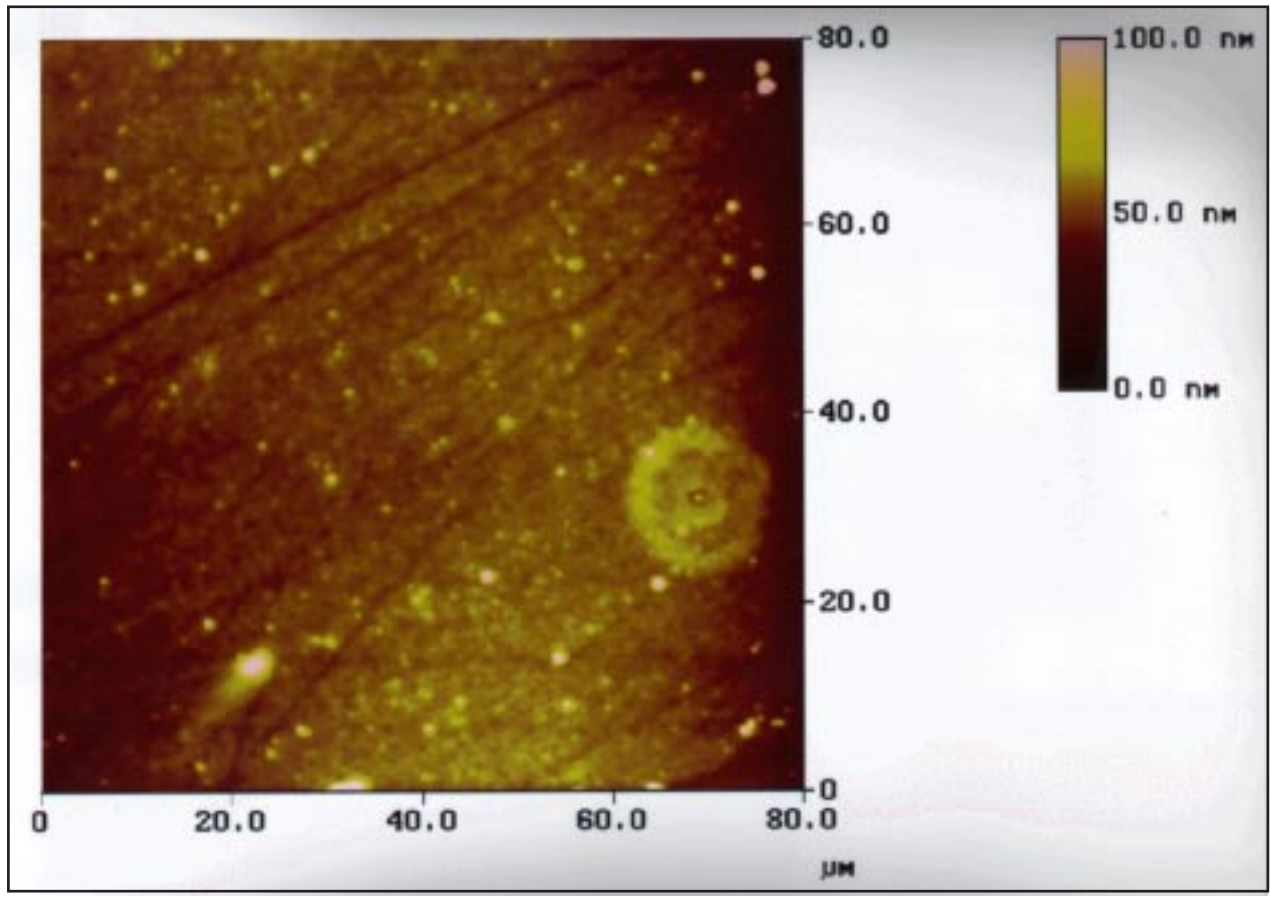

Fig. 5.13: Topography image of Sample 2, applied voltage: $V_{app}$ = 25 V.

The slope of the plane in the SP images is steeper for images acquired at higher applied voltages. Taking the average voltage value of a very thin rectangular portion of the image (1.255 μm x 79.373 μm) close to the left and right side of the image makes it possible to quantify the voltage difference, ΔV. Dividing the voltage difference by the distance between the rectangles' centers ($l$ = 76±2 μm) gives a value for the local electric field, $E_{loc}$. Figures 5.14 and 5.15 show the local electric field as a function of the applied voltage, $V_{app}$, for two measurements performed with Sample 1 and 2.

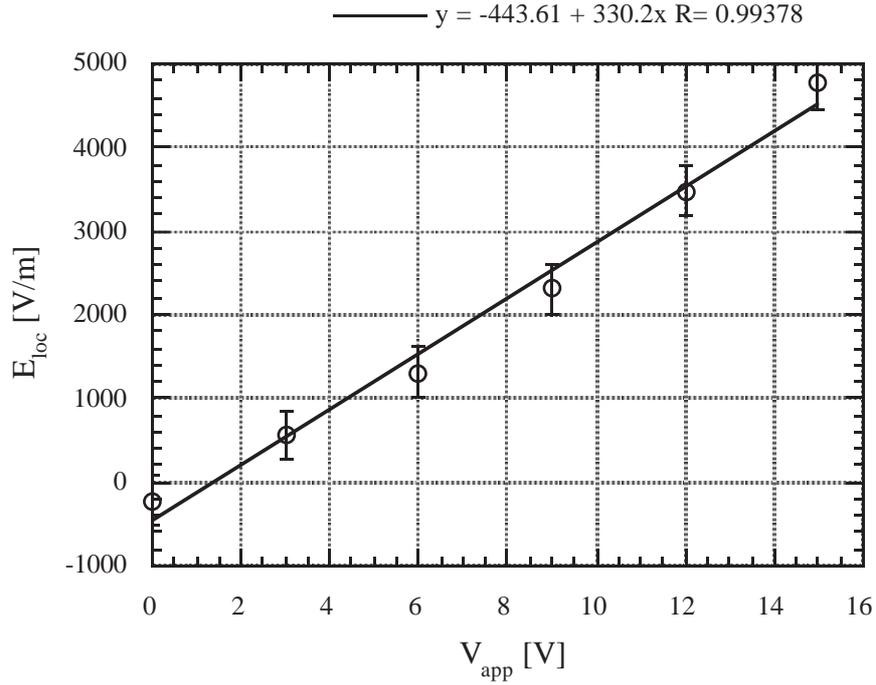

Fig. 5.14: Local electric field, $E_{loc}$, as a function of the applied voltage, $V_{app}$, for Sample 1.

The error bars in the graphs shown in Figure 5.14 and 5.15 were estimated by calculating the absolute error of two data points in each graph, the largest error was taken as the value for the error bars. The error of the SP measurement itself is approximately **8 mV** for the type of cantilever that was used[58]. Since the difference in voltages is calculated by subtracting two voltage values from one another, the absolute error adds up to $\Delta(\Delta V)$ = **16 mV**. The addition of the relative errors of the distance, $\delta(l)$, and the voltage difference, $\Delta V$, gives the relative error of the local electric field, $\delta(E_{loc})$. The largest absolute error calculated this way is $\Delta(\mathbf{E_{loc}})$ = **300 V/m**, which is taken as the value for the error bars. This error seems to be very appropriate for the data and the linear curve fits, except for one data point in Figure 5.15. The SP image from which this data point was calculated shows several scan lines where the measurement was instable, leading to erroneous information. Therefore, this data point was disregarded in the curve fit.

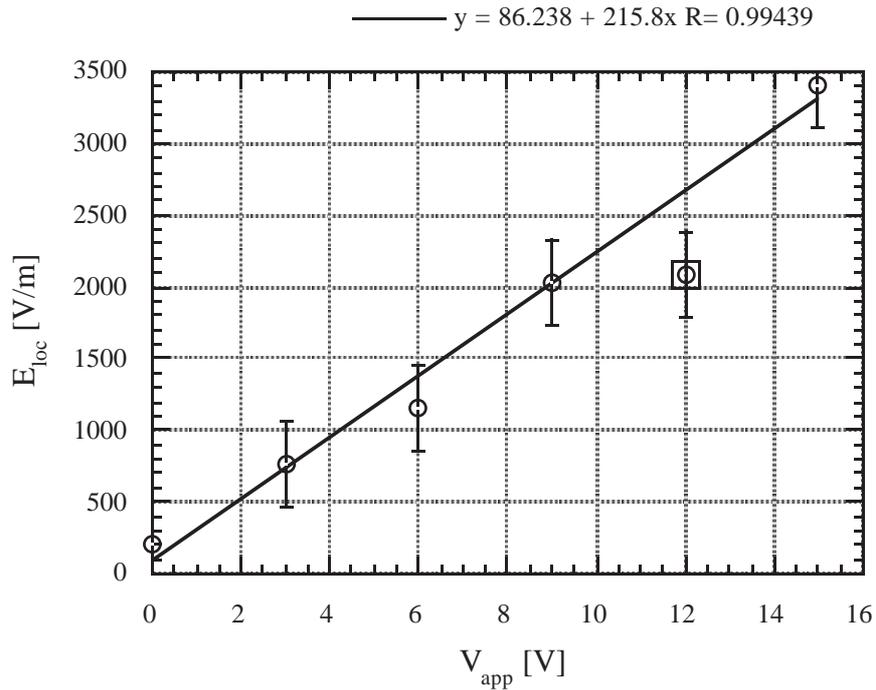

Fig. 5.15: Local electric field, $E_{loc}$, as a function of the applied voltage, $V_{app}$, for Sample 2. The data point for $V_{app}$ = 12 V was disregarded for the line fit.

### 5.6. Summary:

This chapter presented the experimental results, Table 5.1 summarizes the most important electrical and morphological characteristics of the two ZnO samples. The accuracy of the SP measurements depends upon image scale. In small scale SP images, the local morphology of the surface gets convoluted into the SP image, providing a rather large error for the measurement of the difference in local SP between two points on the surface of $\Delta(\Delta U)_{small\ scale}$ = **66 mV**. If on a large scale average values of the SP of complete areas on the surface are taken and compared to each other, the technique is more accurate, providing an error of $\Delta(\Delta U)_{\text{large scale}}$ = **16 mV**.

| Property | Sample 1 | Sample 2 |
|---|---|---|
| Resistivity, $\rho$ [$\Omega \times$cm] | $(2.3 \pm 0.1) \times 10^{-3}$ | $(1.02 \pm 0.07) \times 10^{-2}$ |
| Diameter of elevations, $d_{elev}$ [nm] | 130±40 | 150±30 |
| Height of elevations, $h_{elev}$ [nm] | 32±13 | 39±19 |
| Diameter of protrusions, $d_{pro}$ [nm] | 1700±1100 | 1300±700 |
| Height of protrusions, $h_{pro}$ [nm] | 280±110 | 170±130 |
| Local electric field, $E_{loc}$ [V/m] for $V_{app}$ = 0 V | -223.68±300 | 197.37±300 |
| Local electric field, $E_{loc}$ [V/m] for $V_{app}$ = 3 V | 565.79±300 | 763.16±300 |
| Local electric field, $E_{loc}$ [V/m] for $V_{app}$ = 6 V | 1302.6±300 | 1157.9±300 |
| Local electric field, $E_{loc}$ [V/m] for $V_{app}$ = 9 V | 2315.8±300 | 2026.3±300 |
| Local electric field, $E_{loc}$ [V/m] for $V_{app}$ = 12 V | 3473.7±300 | 2078.9±300 |
| Local electric field, $E_{loc}$ [V/m] for $V_{app}$ = 15 V | 4763.2±300 | 3407.9±300 |

Tab. 5.1: Summary of the most important results.

## *Chapter 6: Discussion*

*This is my theory, and please bear in mind when you read it that it accomodates all the elements involved. And this accomodation used to be the measure of the elegance of theories... before the word "science" came to mean what it means today.*

<div style="text-align: right;">

Lestat de Lioncourt from "Tale of the Body Thief"
by Anne Rice
Chapter 2

</div>

### **6.1. Introduction:**

In this chapter the experimental results are discussed and compared to the samples' characterization by V. Srikant, V. Sergo, and D. Clarke. Furthermore, an explanation is given for the observation that an inverse of the surface topography is convoluted into the SP image. Finally, the observed electric fields in the samples are evaluated and used to make a statement on the quality of the electrical contacts on the samples.

### **6.2. Grain Size:**

Srikant et al. predict a grain size of 100 to 500 nm for both samples. This assumption seems to be supported by the morphology of the samples, which is dominated by elevations. The average diameters of the elevations are $d_{elev}^{Sam\ 1} = (130 \pm 40)$ nm for Sample 1 and $d_{elev}^{Sam\ 2} = (150 \pm 30)$ nm for Sample 2. If each elevation is interpreted as a single grain, then those diameters are actually the grain size, being in accordance with the predicted values. A morphology that shows every single grain as an elevation is not unexpected, since the samples were grown at elevated temperatures ($T_{growth}^{Sam\ 1} = 600$ °C for Sample 1 and $T_{growth}^{Sam\ 2} = 750$ °C for Sample 2). It is known that elevated temperatures can lead to the formation of thermal grooves on the surface of oxides, due to a process of thermal etching at the grain boundaries[59,60]. Such a mechanism would lead to the formation of elevations on the surface, each elevation corresponding to a grain. This idea seems to be supported by the observation that the elevations are more profound on Sample 2, which was grown at a higher temperature than Sample 1.

### **6.3. Electrical Properties:**

The measured resistivities of the two samples ($\rho^{Sam\ 1} = (2.3\pm0.1)\times10^{-3}$ Ω×cm for Sample 1 and $\rho^{Sam\ 2} = (1.02\pm0.07)\times10^{-2}$ Ω×cm for Sample 2) are very low. In general, ZnO does not tend to exhibit its theoretical intrinsic resistivity, which would be more than $10^{10}$ Ω×cm at room temperature, but is rather an n-type semiconductor due to nonstoichiometry in form of zinc excess[61]. The interior of ZnO grains in commercial varistors have resistivities of approximately 1 to 10 Ω×cm[6], which is considered a high conductivity for an oxide. The resistivity of the two samples investigated in this study is roughly three orders of magnitude lower than that value, making their conductivity comparable to that of industrially used n- or p-type doped silicon[62]. This very high conductivity of the ZnO samples is most probably due to the Al doping.

The resistivity values are in agreement with the values for the carrier concentration and the electron mobility determined by V. Srikant et al. The higher resistivity of Sample 2 is in accordance with their hypothesis stating that the samples have potential barriers at the grain boundaries and the height of the barriers is larger in Sample 2.

The I-V curves of the two samples are rather ambiguous. Sample 2 (see Figure 5.2) shows a linear relationship between current and voltage up to an applied voltage of 70 V. Whether or not a breakdown, leading to a nonlinear I-V curve, might occur at higher voltages cannot be concluded. If the sample really shows a breakdown, it should actually be expected to occur at much higher voltages, due to the small grain size. Since the grain size is estimated to be approximately 150 nm and the distance between the contacts is roughly 3 mm, there are around 20,000 grain boundaries between the two contacts. Assuming that no voltage drops at the contacts and in the bulk of the grains, an applied voltage of 70 V would correspond to a voltage drop of 3.5 mV per grain boundary. The common value of the breakdown voltage per grain boundary for commercial ZnO varistors is 3 V[2,3], three orders of magnitude larger than the value calculated above. For this reason, it can not be expected to reach the breakdown voltage with applied voltages as low as 70 V.

The I-V curve of Sample 1 (see Figure 5.1) actually showed a nonlinear behavior for an applied voltage larger than 15 V, trailing off to current values higher than for the linear case. It might be that this actually corresponds to an electrical breakdown, which would indicate varistor behavior. This would be in accordance with the assumption of a low height of the potential barriers at the grain boundaries of Sample 1, enabling a small voltage to produce the breakdown, while Sample 2 has higher potential barriers, so that in this case no breakdown occurs. Nevertheless, the local breakdown voltage in Sample 1 would then be less than 1 mV, which seems unlikely. There is a different explanation for the increase of conductivity at high voltages, which seems more probable: it was observed that the temperature of Sample 1 increased strongly at high applied voltages, due to Joule heating. The conductivity of the ZnO semiconductor can be expected to rise with temperature, due to the thermal excitation of electrons across the band gap from the valence to the conduction band[63,64]. In this case the rise in conductivity would not indicate varistor behavior. As was the case for Sample 2, it can not be concluded whether an electrical breakdown due to a varistor effect might occur at much higher values of the applied voltage.

### 6.4. Effect of the Topography on the Surface Potential Images:

It is a well known and documented fact that tip-related artifacts appear in potentiometry performed with an STM[65]. Not surprisingly, the same seems to be true for SPI performed with an EFM, which is documented by the observation that an inverse of the topography of the surface is convoluted into the surface potential images. This effect can be explained qualitatively with a rather simple model. If the sample was absolutely flat, the tip would have the same distance to its surroundings while scanning the surface (see Figure 6.1). In this case only changes of the electrostatic force would contribute to the surface potential image.

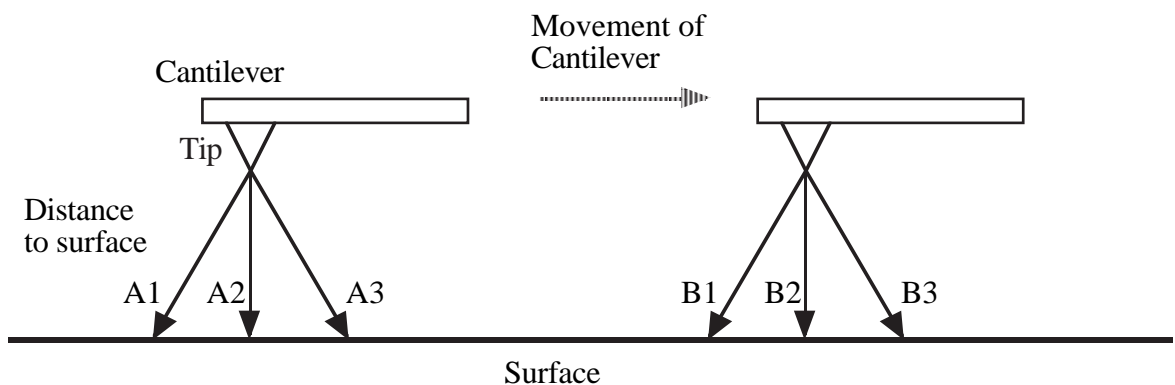

Fig. 6.1: Tip scanning a perfectly flat surface. Distances to the surface stay the same: A1 = B1, A2 = B2, and A3 = B3.

If the surface is not perfectly flat, the technique of EFM still keeps the distance to the surface at a fixed value. Nevertheless, this technique could only give an accurate image of the electrostatic force if only one single atom on the surface, the one which is closest to the tip, exerted a force on it. This is not true, since the surroundings of this atom also exert a force on the tip, so that the tip feels a cumulative effect. In the case of the tip tracing an elevation, the surroundings of the closest atom are farther away than on a flat part of the surface (see Figure 6.2). Since the electrostatic force decays with distance, the cumulative force felt by the tip is weaker than in the flat case.

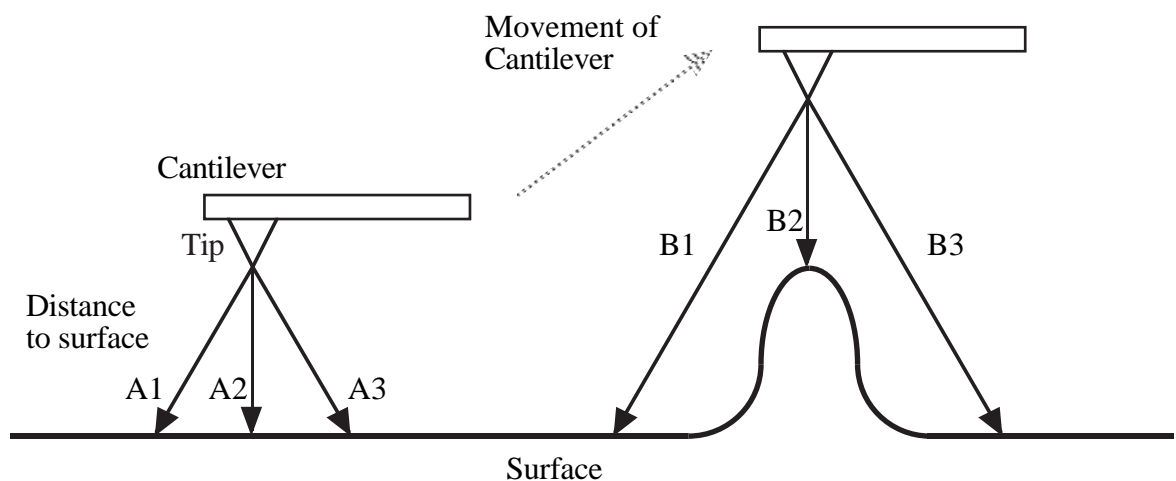

Fig. 6.2: Tip scanning an elevation. Only the distance to the closest atom stays the same: A2 = B2. The distance to the surroundings increases: B1 > A1 and B3 > A3.

In the case of the tip tracing a deepening, the surroundings of the closest atom are closer to it than on a flat part of the surface (see Figure 6.3), and the cumulative force felt by the tip is stronger than in the flat case. Therefore, when the tip traces features on the surface, the force on the tip is going to change, even if the electric field in the sample is constant. The force becomes weaker on elevations and stronger in deepenings, leading to a negative image of the topography.

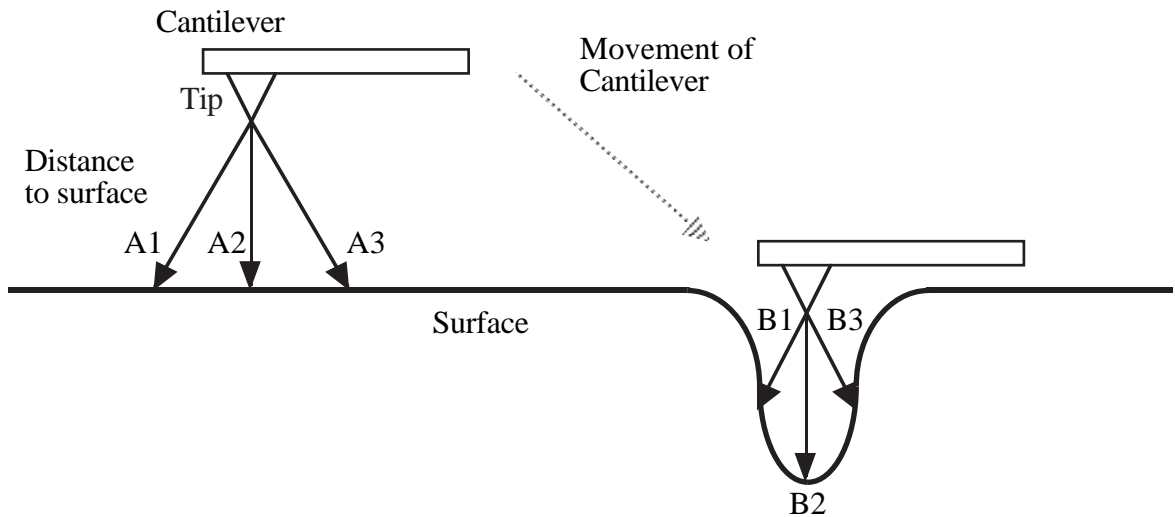

Fig. 6.3: Tip scanning a deepening. Only the distance to the closest atom stays the same: A2 = B2. The distance to the surroundings decreases: B1 < A1 and B3 < A3.

### 6.5. Electric Field:

It was discussed above that the voltage drop per grain boundary is expected to be very small, in the order of 1 to 3.5 mV. Since the error due to the effect of the topography on the SP image was estimated to be approximately 66 mV, a direct observation of single grain boundaries would not provide valuable data. Instead, large sections of the samples were imaged, showing a slope in the surface potential, corresponding to the gradual change in the electric field. The local electric field could be measured this way for several different voltages and was plotted as a function of the applied voltage (see Figures 5.14 and 5.15). The plots revealed a linear relationship between the applied voltage and the local electric field, as is to be expected. Using the linear curve fits of the data, more accurate values of $E_{loc}^{calc}$, the local electric field, can be calculated. The distance, D, between the contacts that were used to apply the voltage to the samples were $D^{Sam\ 1} = 3.3$ mm for Sample 1 and $D^{Sam\ 2} = 3.2$ mm for Sample 2. Multiplication of D and $E_{loc}^{calc}$ gives the total voltage that dropped across the samples, $V_{drop}$. For the case of applied voltage $V_{app}$ = 15 V the calculated values of the local electric field are $E_{loc}^{calc\ /\ Sam\ 1}$ = 4510±300 V/m for Sample 1 and $E_{loc}^{calc\ /\ Sam\ 2}$ = 3320±300 V/m for Sample 2 (the value for the error is the same one that was used for the plots of the local electric field). This gives values for the voltage dropped across the samples of $V_{drop}^{Sam\ 1}$ = (14.9±1.0) V for Sample 1 and $V_{drop}^{Sam\ 2}$ = (10.6±1.0) V for Sample 2. The difference between $V_{drop}$ and $V_{app}$ is the voltage that drops at the contacts, $V_{con}$. For the case of $V_{app}$ = 15 V the values for the drops at the contacts are therefore $V_{con}^{Sam\ 1}$ = (0.1±1.0) V for Sample 1 and $V_{con}^{Sam\ 2}$ = (4.4±1.0) V for Sample 2. Therefore, in Sample 1 only **(0.7±7) %** of the applied voltage are dropped at the contacts, whereas in Sample 2 approximately **(29±7) %** of the applied voltage are dropped at the contacts. This means that the contacts on Sample 1 are better than those on Sample 2.

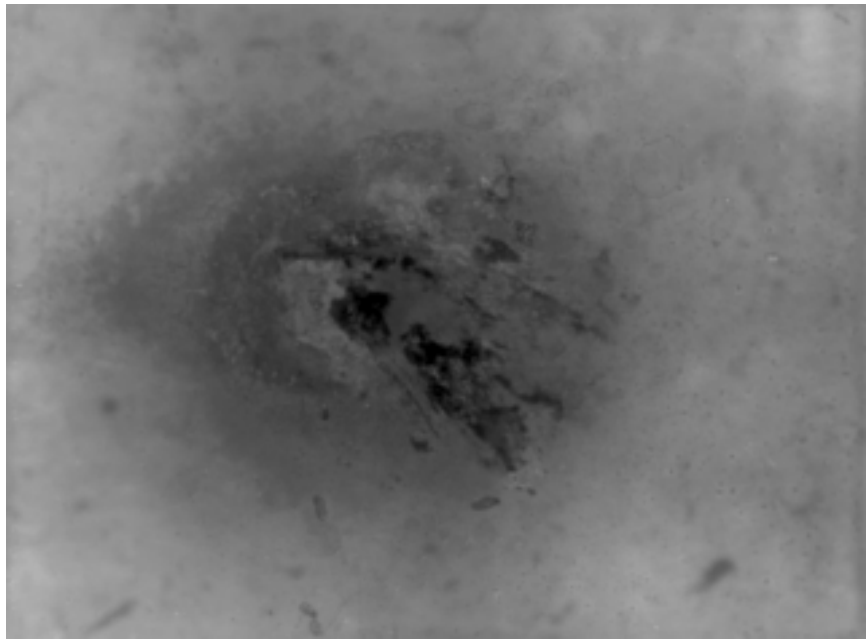

Fig. 6.4: In contact on Sample 1, picture taken from an optical microscope, magnification: 200x.

An observation of the contacts with an optical microscope reveals a possible explanation for this difference: the contacts on Sample 2 (see Figure 6.5) form a clearly visible and defined spot on the surface, whereas the contacts on Sample 1 (see Figure 6.4) form rather a diffuse area. The sintering process that was used to burn the contacts into the surface therefore seems to have been more successful with Sample 1, leading to a diffusion of In into the ZnO and an intimate contact, providing a low resistivity electrical contact. In Sample 2 the In remained as a compact pad on top of the surface and therefore does not supply the necessary intimate junction for a low resistivity contact.

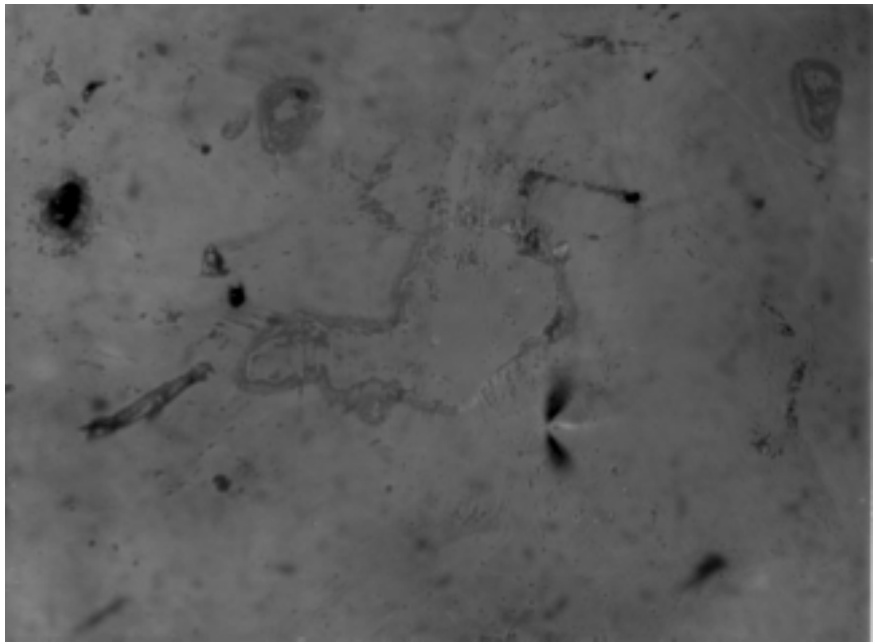

Fig. 6.5: In contact on Sample 2, picture taken from an optical microscope, magnification: 200x.

### 6.6. Summary:

This chapter presented a discussion of the experimental results. The observed elevations in the samples' topography were interpreted as grains, giving an average grain size of (130±40) nm for Sample 1 and (150±30) nm for Sample 2. These values are in accordance with the predictions of V. Srikant, V. Sergo, and D. Clarke. The measured electrical properties of the two samples are also in accordance with the findings of V. Srikant et al., but give no direct evidence for the existence of potential barriers at the grain boundaries. The nonlinear I-V characteristic of Sample 1 is attributed to a rise in temperature of the sample during the measurement.

A simple, qualitative model was given as an explanation for the observation that an inverse of the surface topography is convoluted into the SP images. The effect is attributed to the surroundings of the surface atom closest to the tip, exerting a strong force on the tip when in dips into a deepening and a weak force when it arches over an elevation.

Finally, the observed electric fields in the samples were evaluated. The measurements are very accurate, they even allow to make a statement on the quality of the electrical contacts on the samples by calculating the voltage drop between them. Sample 1 has good contacts, only (0.7±7) % of the applied voltage are dropped at them, whereas the contacts on Sample 2 are not as good, leading to a voltage drop of approximately (29±7) % of the total applied voltage.

*Chapter 7: Conclusion and Outlook*

> *I do not know what I may appear to the world; but to myself I seem to have been only like a boy playing on the sea-shore, and diverting myself in now and then finding a smoother pebble or a prettier shell than ordinary, whilst the great ocean of truth lay all undiscovered before me.*
>
> Isaac Newton
> "Brewster's Memoirs of Newton"
> Volume 2, Chapter 27

**7.1. Conclusion:**

In this study in-situ surface potential measurements were performed with an atomic force microscope, using the technique of Surface Potential Imaging, to investigate laterally applied electric fields in zinc oxide thin film samples for the first time. The local change in electric field across the samples was monitored and quantified. It was observed that an inverse of the morphology of the surface is convoluted into the surface potential image, and the magnitude of this effect was quantified by taking surface potential images without an applied electric field. For the given samples and measurement conditions a **height difference** of **80 nm** in the topography image resulted in a **voltage difference** of approximately **66 mV** in the surface potential image. A simple model provided in this study attributes this observation to the surroundings of the surface atom closest to the imaging tip. When the tip descends into a deepening, the surroundings of the closest surface atom exert a strong force on it, when the tip ascends above an elevation, they exert a weak force.

It is assumed that the ZnO samples examined in this study have potential barriers at the grain boundaries so that a large part of an applied voltage should drop in the region close to the grain boundaries. Since the estimated **grain size** of the samples is very small, approximately **140 nm**, the voltage drop per grain boundary is minute for any reasonably high applied voltage, in the order of 1 to 3.5 mV. This is more than an order of magnitude smaller than the error introduced by the topography, and therefore the observation of single grain boundaries was not performed in this study.

The surface potential images taken in this study for quantitative analysis were on a large scale, 80 μm, and complete areas of the surface were evaluated for a mean value of the voltage close to the sides of the image. In doing this, the effect of the topography was averaged out and the voltage drop across the 80 μm sector could be measured within the **accuracy of the technique**, which is approximately **16 mV** for the cantilever that was used. The electric field was calculated from this voltage drop and compared to the applied voltage, making it possible to determine the voltage drop at the contacts, which is a measure of the quality of the electrical contacts.

### 7.2. Outlook:

In the future, the technique of SPI measurements with the AFM could be used as a very valuable tool to examine the change of an applied electric field at grain boundaries. To measure the effect at a single grain boundary, samples are needed with a much larger grain size than that of the ones used in this study. Bicrystals or a polycrystalline sample with a grain size large enough to apply microcontacts would be perfect, so that a large voltage can be dropped across a single boundary. It has been shown that with these set-ups I-V curves of single grain boundaries can be measured, but the interpretation of these is somewhat difficult, since effects taking place in the sample and at the contacts are convoluted[17,21,22,24,25]. A combination of these single-grain-boundary I-V curves with an in-situ measurement of the electrical field with the SPI technique could be used to obtain very precise values for the breakdown-voltage of single grain boundaries in ZnO varistors, an issue that has been often discussed in literature and is still not resolved.

## *References:*

*Books must follow sciences, and not sciences books.*

Francis Bacon
"Proposition touching Amendment of Laws"